\documentclass[12pt]{article}
\usepackage[top=0.75in, bottom=0.75in, left=0.9in, right=0.9in]{geometry}
\newcommand{\vect}[1]{\mbox{\boldmath $ #1$}}
\usepackage{color}

\def\bSig\boldsymbol{\Sigma}
\newcommand{\bbeta}{\boldsymbol{\beta}}
\newcommand{\bX}{\mathbf{X}}
\newcommand{\btheta}{\boldsymbol{\theta}}

\usepackage[figuresright]{rotating}

\newcommand{\single}{\baselineskip 15pt}

\newcommand{\ind}[1]{\mathbb{I}\lbrace#1\rbrace}

\usepackage{graphicx}
\usepackage{color}
\usepackage{hyperref}
\usepackage{makeidx}
\usepackage{amsmath,amsfonts}
\usepackage{caption}
\usepackage{pdfpages}
\usepackage{chngcntr}      
\usepackage{booktabs}
\usepackage{subcaption}
\usepackage[comma]{natbib}
\usepackage{tikz}
\usetikzlibrary{arrows,decorations.pathreplacing,calc,datavisualization.formats.functions}
\usepackage{pgfcore}
\tikzset{
    >=stealth',
    shorten <>/.style={ shorten >=#1, shorten <=#1} 
}

\makeindex
\usepackage{epsfig}

\usepackage[deletedmarkup=sout]{changes}
\definechangesauthor[name={Jon Fintzi}, color={red}]{jf}
\definechangesauthor[name={Dean Follmann}, color={blue}]{df}

\makeatletter
\@namedef{Changes@AuthorColor}{red}
\colorlet{Changes@Color}{red}
\makeatother

\begin{document}

\label{firstpage}

\begin{center}
	{\LARGE Assessing Vaccine Durability in Randomized Trials Following Placebo Crossover}\\   
	\ \\
	\large{Jonathan Fintzi and 
		Dean Follmann,\\}
	{\small
		Biostatistics Research Branch, National Institute of Allergy and Infectious Diseases\\
		Rockville, Maryland, U.S.A.\\
		email addresses: \url{jon.fintzi@nih.gov} and \url{dean.follmann@nih.gov}}
\end{center}
%
%

\renewcommand{\abstractname}{Abstract}
\begin{abstract}
Randomized vaccine trials are used to assess vaccine efficacy and to characterize the durability of vaccine induced protection.  If efficacy is demonstrated, the treatment of placebo volunteers becomes an issue. For COVID-19 vaccine trials, there is   broad consensus that placebo volunteers should be offered a vaccine once efficacy has been established.
This will likely lead to most placebo volunteers crossing over to the vaccine arm, thus complicating the assessment of long term durability. We show how to analyze durability following placebo crossover and demonstrate that the vaccine efficacy profile that would be observed in a placebo controlled trial is recoverable in a trial with placebo crossover. This result holds no matter when the crossover occurs and with no assumptions about the form of the efficacy profile. We only require that the vaccine efficacy profile applies to the newly vaccinated irrespective of the timing of vaccination. We develop different methods to estimate efficacy within the context of a proportional hazards regression model and explore via simulation the implications of placebo crossover for estimation of vaccine efficacy under different efficacy dynamics and study designs. We apply our methods to simulated COVID-19 vaccine trials with durable and  waning vaccine efficacy and a total follow-up of two years.
\end{abstract}

\noindent%

\newpage
\section{Introduction}
\label{s:intro}
Randomized phase III clinical trials are used to definitively demonstrate the efficacy of candidate vaccines. Volunteers are randomized to receive vaccine or a placebo and followed for a period of time to assess whether the vaccine reduces the rate of disease acquisition. An important question in vaccine development is whether vaccine induced protection is durable. For COVID-19 vaccines, questions surrounding vaccine durability are important as acquired immunity against seasonal and other coronaviruses ranges from six months to two years \citep{poland2020sars,choe2020waning}.
Clinical trials for vaccines against COVID-19 plan to follow participants for up to two years \citep{modernaprotocol}. 

To assess long term safety and durability, long term blinded  follow-up of the original placebo and vaccine arms is ideal \citep{who2020placebo}. From an ethical perspective, placebo volunteers should be offered a vaccine once efficacy is established \citep{wendler2020covid}.
However, vaccination of placebo volunteers may occur before it is known whether vaccine induced protection is durable.  Besides waning of efficacy, there is concern that the vaccine might eventually cause harm, i.e. negative vaccine efficacy,  in subgroups. Such harm is known as vaccine associated enhanced disease (VAED) and has been observed in other contexts, such as the Dengvaxia vaccine in seronegative individuals \citep{sridhar2018effect}.

It might seem that the ability to assess vaccine durability following placebo crossover is completely lost once there is no longer an unvaccinated control group \citep{who2020placebo}. However, at the point of crossover the study remains a randomized trial, albeit of immediate vs deferred vaccination.   This contrast allows  the vaccine efficacy (VE) profile for  a standard non-crossover trial to be  recovered with placebo crossover \citep{follmann2020placros}. The only additional assumption that is required is that the same VE profile applies to the newly vaccinated irrespective of the timing of vaccination e.g. June or December. 

Crossover trials for absorbing endpoints, such as infection or death, have been discussed in the literature \citep{nason2010,makubate2010planning}. However, these methods apply to estimation of a intervention effect that stops once the intervention is removed. Vaccination is different as the benefit lingers and our goal is to see how the intervention effect varies with time. Crossover has been discussed for vaccine trials, but only for the placebo arm and only to measure immune response \citep{follmann2006augmented}. Delayed vaccination has been used in an Ebola vaccine trial, but to serve as control group prior to deferred vaccination \citep{henao2017efficacy}.


We establish that vaccine durability can be accurately assessed following placebo crossover under fairly mild assumptions.  We demonstrate how to estimate VE as a function of time since vaccination
under placebo crossover using proportional hazards (PH) regression \citep{cox1972regression} where VE, defined as as 1 minus the hazard ratio (HR), is allowed to depend on time through use of time-varying covariates \citep{therneau2013modeling}. We specify log-linear and P-spline functions to allow for a variety of shapes for the VE profile
and  provide a justification for using calendar time as the natural timescale in vaccine trials where risk can vary substantially  with calendar time. 
We discuss and evaluate different approaches for crossing over placebo participants, and estimation is affected by the timing and pace of crossover, and unobservable heterogeneity in risk. We evaluate our methods by simulation and analyze two simulated COVID-19 vaccine trials that vaccinate placebo volunteers after efficacy is established.   

\section{Vaccine Efficacy Under Placebo Crossover}
\label{sec:gen_development}

\subsection{Conceptual Development}
\label{subsec:concepts}

Consider a vaccine trial where volunteers are randomized to receive vaccine or placebo. For now, assume that everyone is enrolled at the same time, so calendar time and time since randomization are aligned. All participants are followed over the period $[0,\ 2 \tau]$, and a blinded crossover occurs at time $\tau$, at which point the volunteers randomized to vaccine receive placebo, and the volunteers randomized to placebo receive vaccine. Following crossover, both arms are vaccinated and thus comparative efficacy might seem lost as there is no control group. However, a randomized trial remains, though now as a trial of immediate vs deferred vaccination; these assignments correspond to the original vaccine and placebo arms. This 'rebranded' randomized trial can still provide information about vaccine durability, even after the point of crossover. 

To illustrate, suppose we have case counts for the two randomization arms over the two periods, $(0,\ \tau]$ and $(\tau,\ 2\tau]$. Suppose that the vaccine:placebo case split is 20:100 in period one, and in period two we observe a case split of 20:12 in the original vaccine arm:deferred vaccination arm. Using a person-time analysis, and imagining the denominators are so large that they cancel out, we obtain a simple estimate of the period one VE as $\widehat{VE}_1 = 1-20/100 = 0.80$. Assume now that this VE applies to the newly vaccinated participants in the second period with 12 cases.  With this assumption we can estimate $N_2^{plac}$, the number of cases for a counterfactual placebo group in period two,
as we are assuming $0.80 = 1 - 12/\hat{N_2^{plac}}$, which yields $N_2^{plac}=12/.2=60.$
We then contrast the counterfactual placebo case count of 60 to the 20 observed cases in the original vaccine arm to obtain an estimate of placebo controlled VE in period two, $\widehat{VE}_2 = 1 - 20/60=0.667$. Based on these crude estimates, we conclude that VE has waned as efficacy has dropped from 80\% in period one to 66.7\% in period two.  

The crux of this example is that the VE among the newly vaccinated is portable across periods. That is, the original placebo arm receives the same immediate benefit from vaccination that the original vaccine group received, regardless of changes in the population attack rate. Additional considerations for a period focused approach are discussed in Follmann, et al. (2020)\cite{follmann2020placros}. 
While a period focused approach is simple and clear, a more natural development allows VE to vary smoothly with time. The Cox proportional hazards (PH) model allows this to be easily accomplished. Under the PH model, the baseline placebo hazard function for the time to disease is arbitrary  and the effect of vaccination
induces a hazard proportional to the baseline hazard. We can formulate the hazard function corresponding to the previous example for a standard trial with no crossover as
\begin{equation}
	\label{eqn:model1}
	h(t) = h_0(t) \exp\{ Z \ind{t<\tau}  \theta_1 + Z \ind{t \geq \tau} \theta_2 \},
\end{equation}
where $t$ is time since randomization, $h_0(t)$ is the placebo hazard function, and $Z$ is the vaccine assignment indicator. Vaccine efficacy is defined as the relative change in the instantaneous risk of acquiring disease, and is given by  $1-\exp(\theta_1)$ for period one and $1-\exp(\theta_2)$ for period two.   

Suppose placebo volunteers are vaccinated at the end of period one.   
Assuming the vaccine effect for the newly vaccinated applies in period two, we can write
\begin{eqnarray*}
	h(t)&=& h_0(t) \exp\{ Z \ind{t<\tau}  \theta_1 + Z \ind{ t \geq \tau} \theta_2 + (1-Z)\ind{t \geq \tau}  \theta_1 \} \\
	&=& \lambda_0(t) \exp\{ Z \ind{t<\tau}  \theta_1 + Z \ind{ t \geq \tau} \theta_2 \},
	\label{eqn:model2}
\end{eqnarray*} 
\noindent where the baseline hazard $\lambda_0(t) = h_0(t) \exp\{ \ind{t\geq \tau} \theta_1 \}$
applies to the original placebo arm for both periods.
The first line parameterizes the placebo controlled 
VE for the newly vaccinated in period two 
as $1-\exp(\theta_1)$ and the VE for the original vaccinees in period two as $1-\exp(\theta_2)$.  
Analogous to the simple example where we recovered
the placebo case count $N_2^{plac}$ after time $\tau$,
here we  can  recover the counterfactual placebo hazard function $h_0(t)$ 
for $t \geq \tau$ as $h_0(t) = \lambda_0(t) \exp(-\theta_1)$.

Equation (\ref{eqn:model1}) is deceptively simple.  
Generalizations of this idea allow us to recover an arbitrary placebo controlled 
VE curve long after the placebo group has been completely vaccinated
and with no assumptions about the baseline hazard function for an actual or counterfactual
placebo arm. 
To  illustrate,  suppose that the placebo controlled hazard
is given by a log-linear function of time.
\begin{equation}
	h(t) = h_0(t) \exp\{ Z (\theta_1 + \theta_2 t)\}
	\label{eqn:loglin_decay_haz}
\end{equation}
If crossover to vaccine occurs at time  $\tau$ in the placebo arm, the resultant 
hazard is 
\begin{eqnarray}
	h(t)& =& h_0(t) \exp\left\{ Z (\theta_1 + \theta_2 t) + (1 - Z) \ind{t\geq\tau}(\theta_1 + \theta_2 (t - \tau)) \right\}. \nonumber \\
	&=& \lambda_0(t) 
	\exp\left\{ Z \theta_2 t + (1 - Z) \ind{t\geq\tau} \theta_2 (t - \tau) \right\}.
	\label{eqn:log-lin-cross}
\end{eqnarray}
\noindent where $\lambda_0(t)=h_0(t)\exp( \ind {t\geq\tau} \theta_1)$.  
A visualization of this hazard function is given in Figure 1.

To better understand what happens in terms of estimation, suppose that an event
occurs in the original placebo arm at time $t=s< \tau$,
which is post randomization, but before crossover.
This scenario is illustrated in Figure 1. The partial likelihood contribution  for this event under model
(\ref{eqn:log-lin-cross}) reduces to 
\begin{equation*}
	\frac{1}{\sum_{i \in R(s)} \exp\{ Z_i(\theta_1 + \theta_2 s) \}}
\end{equation*}
\noindent where $R(s)$ is the set of indices for volunteers
who remain event free at time $s$.
Thus events  prior to $\tau$ allow estimation of $\theta_2$ and, crucially, $\theta_1$.    
Next suppose an event occurs at time $t=\tau +s$ post randomization 
to an individual in the original placebo arm who was vaccinated
at time $\tau$ and 
who now has been vaccinated for $s$ days (Figure 1).    The partial likelihood contribution for this 
event is  
\begin{equation*}
	\frac{\exp(\theta_2 s)}{\sum_{i \in R(\tau+s)} \exp\{ Z_i (\tau+s) \theta_2 +(1-Z_i) s \theta_2 \} }.
\end{equation*} 
We see that $\theta_1$ is gone as the baseline hazard, $\lambda_0(\tau+s) = h_0(\tau+s)\exp(\theta_1)$, cancels out of the numerator and denominator.  
Thus the pre-crossover period 
completely  determines the reliability of the estimate of $\theta_1$.
As a result, longer pre-crossover 
periods with more events are desirable to better estimate $\theta_1$
Additionally, as $\tau$ approaches zero,
the covariate value for the original vaccine arm $(\tau+s)$  is very close to the 
covariate value for the original placebo arm $s$. Insufficient variation in covariate values makes estimation of the associated regression slope more difficult. Thus the benefit of a longer pre-crossover period persists in estimation of $\theta_2$.
in addition to helping to estimate $\theta_1$.   
We note that there is nothing really special about this kind of cancellation.  Consider a Cox regression with two covariates; a treatment indicator and a sex indicator.  Even if all the women eventually dropout or have events, we continue to accrue information about the effect of treatment, provided we still have men at risk.  



\begin{figure}
	\centering
	\includegraphics{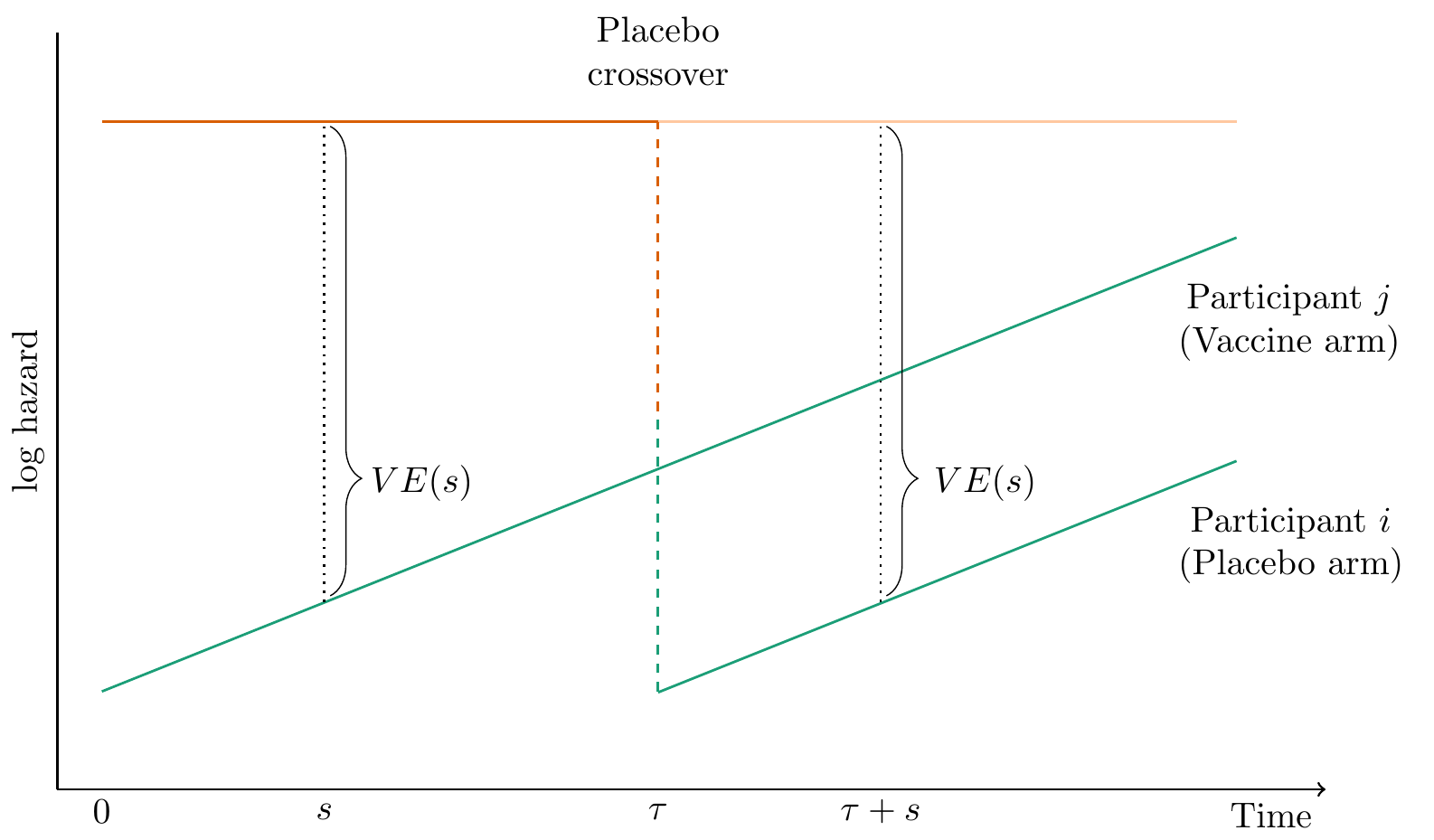}
	\caption{Log hazard for two study participants: $i$, who is initially given  placebo (orange line), and $j$ who is initially given vaccine (green line). Vaccine efficacy wanes (i.e. log hazard increases) as a function of time since vaccination. At time $\tau$, participant $i$ is given vaccine and follows the same efficacy profile as participant $j$. The 'baseline' hazard function for original placebo  participant $i$  is $\lambda_0(t) = h_0(t)$ prior to crossover and $\lambda_0(t) = h_0(t) \exp\{\theta_1 + \theta_2 \left(t - \tau \right)\}$ after crossover.
		In this figure, $h_0(t)$ is constant, but this is not required.}
	\label{fig:crossover_diagram}
\end{figure}

\subsection{General Development}
\label{subsec:gen_setup}
We now develop this approach for the more realistic setting of a staggered entry trial and consider more general models for VE over time. Let $t\geq 0$ now index time since study initiation, not time since randomization. For each subject, $i,\ i\in 1,\dots,N$, the data, $\left(\tau_i^{(e)},\tau_i^{(v)},T_i,C_i,Z_i,\bX_i\right)$, consist of the times of study entry and vaccination, $ \tau_i^{(e)}$ and $\tau_i^{(v)}$, with $\tau_i^{(v)}\geq\tau_i^{(e)}>0,$, the time to symptomatic COVID-19 or end-of-followup, $T_i = \min(Y_i, C_i)$ with $Y_i$ the true but possibly unobserved event time, treatment assignment $Z_i$, and baseline covariates $\bX_i$. By convention, we take $\tau_i^{(v)}$ to be greater than the study duration if disease occurs prior to vaccination. We also define a time-dependent vaccination indicator, $Z_i(t) = \ind{t>\tau_i^{(v)}}$.

The hazard for participant $i$ is  
\begin{align}
	\label{eqn:general_haz}
	h_i(t) &= \left\lbrace \begin{array}{ll}
		0 &,\ t\leq\tau_i^{(e)}, \\
		h_0(t)\exp\left[Z_i(t)f\left(t-\tau_i^{(v)}; \btheta\right) + \vect X' \bbeta \right] &,\ t > \tau_i^{(e)},
	\end{array} \right .
\end{align}
where $h_0(t)$ is the arbitrary `reference' hazard for a placebo group,  $\btheta$  a vector of parameters governing VE over time, $\vect X$ a vector of baseline covariates
and $\bbeta$ a vector of parameters.  The hazard is zero before an individual enters the trial by convention as the hazard function applies to volunteers who are uninfected when randomized.  
We calculate VE at time $s$ post--vaccination as one minus the ratio of vaccine to placebo hazards, i.e., 
\begin{equation*}
	VE(s) = 1 - \exp[ f(s ;\btheta)].   
\end{equation*}

The model, (\ref{eqn:general_haz}), encompasses standard trials with parallel arms in which case
$\tau_i^{(v)} = \infty$ for placebo volunteers  and placebo-crossover trials in 
which case $\tau_i^{(v)} > \tau_i^{(e)}$ for participants on the placebo arm.  Following crossover of all placebo subjects, $Z_i(t)=1$ for all study participants, hence the hazard ratio for any pair of subjects with the same $\vect X$  is only depends on the contrast in their times since vaccination:
\begin{equation*}
	\frac{h_i(t)}{h_j(t)} = \frac{\exp\{ f(t-\tau_i^{(v)};\btheta)\}}{\exp\{ f(t-\tau_j^{(v)};\btheta)\}}.
\end{equation*}
This ratio is 1 for a constant VE model and so following crossover, 
there is no additional information about a constant VE, just as there
is no additional information about $\theta_1$ in the log-linear
decay model.   This differs from the standard parallel arms trial
where such information accrues throughout follow-up.

\begin{figure}[htbp]
	\centering
	\begin{subfigure}[b]{0.49\textwidth}
		\label{a}
		\includegraphics{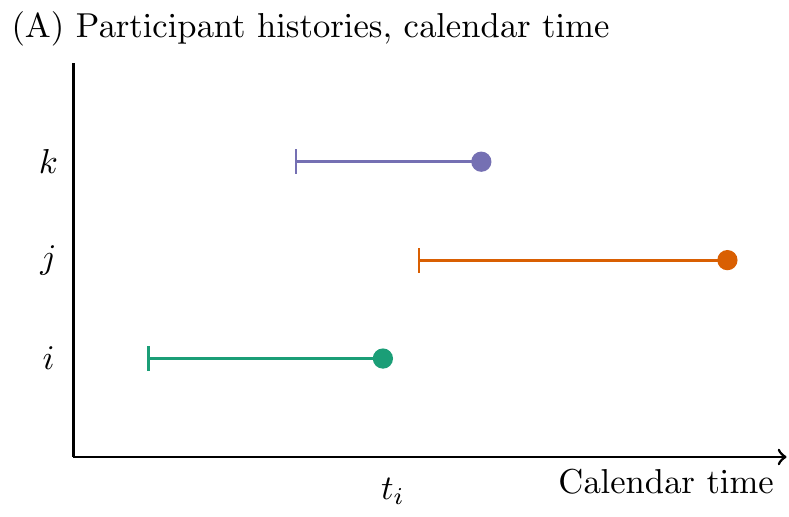}
	\end{subfigure}
	\hfill
	\begin{subfigure}[b]{0.49\textwidth}
		\label{b}
		\includegraphics{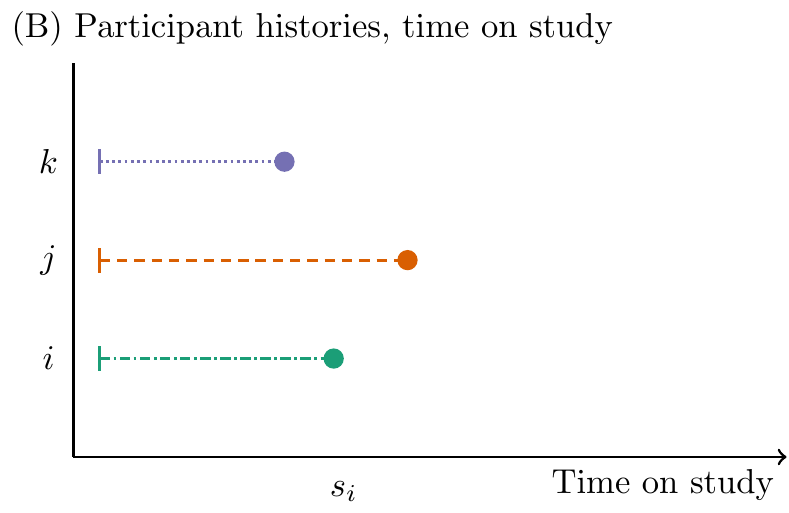}
	\end{subfigure}
	\\\bigskip\vspace{0.1in}
	\begin{subfigure}[b]{0.49\textwidth}
		\label{c}
		\includegraphics{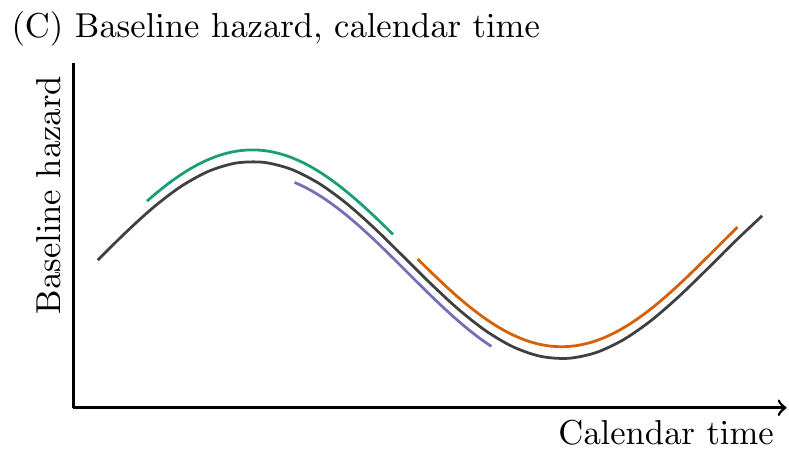}
	\end{subfigure}
	\hfill
	\begin{subfigure}[b]{0.49\textwidth}
		\label{d}
		\includegraphics{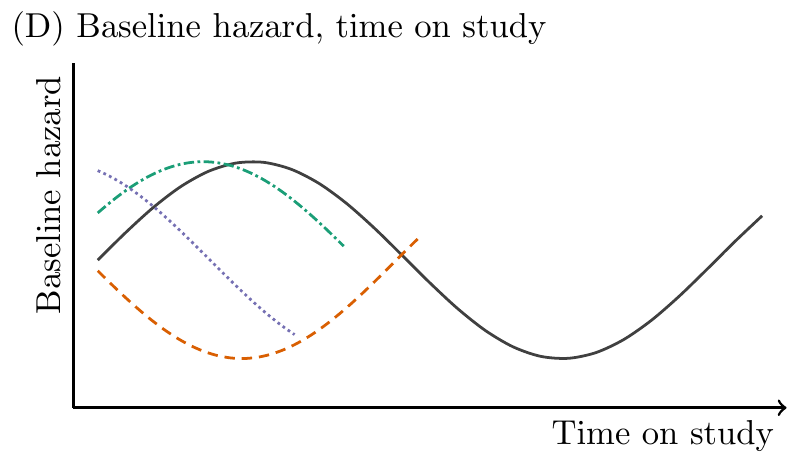}
	\end{subfigure}
	\caption{Participant histories and baseline hazards when the data are indexed in calendar time or aligned on times of study entry. The true data generating mechanism is indexed in calendar time. (A) vs. (B): Aligning the data on study entry changes the risk set as $k$ falls out of the risk set at $i$'s event time and $j$ is incorrectly introduced into the risk set. (C) vs. (D): Baseline hazards are no longer proportional after the data are aligned on study entry.}
	\label{fig:caltime_vs_studytime}
\end{figure}

While it is standard to have the time index for the Cox model be time since randomization, calendar time is  a more natural index for the Cox model as specified in (\ref{eqn:general_haz}) for trials where risk can wax and wane over time. In our setting, aligning the data on study entry induces non-proportional hazards and distorts the risk set in the Cox partial likelihood at each event time. These phenomena are diagrammed in Figure \ref{fig:caltime_vs_studytime}. 


Suppose participant $i$ acquires disease at calendar time $t_i$ after being on study for a period 
$s_i = t_i - \tau_i^{(e)}$ (Panels A and B) and we use 
model (\ref{eqn:general_haz}) with calendar time index (Panel C).   
Setting aside baseline covariates, the partial likelihood contribution at calendar time $t_i $ is
\begin{equation*}
	\frac{h_0(t_i) \exp\{ Z_i(t_i) f(t_i - \tau_i^{(v)}; \theta) \}}
	{ \sum_{j \in R(t_i)} h_0(t_i) \exp\{ Z_j(t_i) f(t_i - \tau_j^{(v)}; \theta) \}}
\end{equation*}
and the baseline hazards cancel out as they should.  

Now, suppose we align participants on their times of study entry. 
The event of person $i$ at calendar time $t_i$ is at study time $s_i$ and the associated
study time risk set is $\widetilde{\mathcal{R}}(s_i)$.
The calendar time for participant $i$ is $t_i = s_i + \tau_i^{(e)}$, whereas
for a generic participant $j$ it is a different calendar time $t_j = s_i + \tau_j^{(e)} $.  
Since the hazard truly depends on calendar time, the partial likelihood contribution at study time $s_i$ is
\begin{equation*}
	\frac{h_0(\tau_i^e + s_i) \exp\{ Z_i(\tau_i^e+s_i) f(s^*_i(s_i); \theta) \}}
	{ \sum_{j \in  \tilde{R} (s_i)} h_0(\tau_j^e +s_i) \exp\{ Z_j(\tau_j^e+s_j) f(s^*_j(s_i); \theta)\} }
\end{equation*}
where $s^*_j(s_i)$ is the time since vaccination at study time $s_i$ for person $j$.  
With alignment on time since study entry, the  baseline hazards do not cancel out 
and a partial likelihood contribution which assumes they do will be misspecified.

\subsection{Flexible Models for VE(s)}
\label{subsec:est_ve}

%

The log-linear and piecewise-constant forms of VE(s) discussed above 
are simple and useful to understand behavior of the model
and estimation.    However, the form of waning vaccine 
efficacy can be hard to anticipate for new vaccines and high constant efficacy followed by a quick or smooth decay is possible as are other shapes.
It is thus appealing to  model $f(\cdot)$ semi-parametrically e.g. using penalized cubic P-splines \citep{eilers1996flexible,wood2017generalized,perperoglou2019review}.
Let $\mathbf{P}_L(t;\mathbf{k},\delta)$ denote a P-spline basis of degree $\delta=3$ with $L$ basis terms 
and vector of knot locations $\mathbf{k}$, and let $\boldsymbol{\gamma}$ be a vector of coefficients, with $\gamma_0$ reserved for the log-hazard ratio immediately following vaccination. The hazard for participant $i$ is
\begin{align}
	\label{eqn:pspline_haz}
	h_i(t) &= \left\lbrace \begin{array}{ll}
		0 &,\ t\leq\tau_i^{(e)}, \\
		h_0(t)\exp\left(Z_i(t)\{ \gamma_0 + \sum_{\ell=1}^L\gamma_\ell P_\ell(t - \tau_i^{(v)};\mathbf{k},\delta) \} \right) &,\ t>\tau_i^{(e)}.
	\end{array} \right .
\end{align}
In practice, we center the decay component estimated by the P-spline at zero to ensure identifiability of $\gamma_0$. Note that we need to evaluate the hazard for each participant at every event time, not merely at the time when a person experiences their own event \citep{therneau2017using}. 

Splines can be implemented in the \texttt{SAS} procedure \texttt{PROC PHREG} 
and in \texttt{R}
using the  \texttt{survival} package \citep{survival-package}.  The latter  provides users with a convenient summary method for the linear and non--linear spline effects, which is useful for testing for non-linearity in the decay profile.

\subsection{Crossing Over}
\label{subsec:black}
Our development up to now has implicitly started counting cases
immediately after the first dose of vaccine. In practice vaccine trials often use a per-protocol primary analysis that forgoes  counting disease cases until after the immunization schedule is complete, e.g. seven days following the last dose.   Such an analysis better evaluates the full benefit of immunization.   For such analyses, we  need to symmetrically avoid counting cases in both arms during the second immunization period even if it is counterfactual, that is if volunteers randomized to vaccine are unblinded 
and not immunized.
To achieve this symmetry, a `blackout' period of length $\Delta$ 
can be defined by the hazard function $h(t) \{ 1 - I(t \in [\tau^{(x)}, \tau^{(x)} +  \Delta] \}$, where $\tau^{(x)}$ is the start of the crossover
or unblinding for an individual, and $\Delta$ the 
time from first dose to when cases are counted.   
The consequence is to define
discontinuous pre and post crossover risk intervals for volunteers who complete crossover without having an event.    Volunteers who 
have an event before or during crossover have a single risk interval which ends
in an event or censoring, respectively.   

Placebo crossover might happen in a blinded or unblinded (open label) manner. Blinded crossover is preferred to avoid potentially differential risk behavior as the recently unblinded volunteers originally randomized to vaccine, who now know they are protected, might forgo risk avoidance behavior \cite{follmann2020placros}. 
With open label crossover such differential behavior could cause a spurious waning efficacy in the period immediately following unblinding. One complex approach  to address potential bias from unblinded crossover would be to use covariate adjustment and stratification. Let $\vect X$ denote a vector of covariates measured at baseline, or pre or post unblinding that  predict  risk behavior.  While clinical trials typically avoid use of  post-baseline variables for adjustments, in the open label setting such adjustment may ameliorate bias. Once an individual is unblinded, a new hazard function applies. We illustrate using  log-linear decay:  
\begin{equation}
	h^*(t) = \lambda^*_0(t) \exp\left\lbrace \theta_2 (t-\tau^{(v)})  + 
	\vect X' \bbeta\right\rbrace,
	\label{eqn:unblindedhazard}
\end{equation}
\noindent where $\lambda^*_0(t)$ is the new baseline hazard for the original placebo arm in this new open label milieu. Because crossover of all subjects cannot occur at the same time, there will be a crossover interlude during which the placebo volunteers become vaccinated.    
Thus, at calendar time $t$ during the crossover interlude,
the expanding unblinded cohort would use hazard $h^*(t)$ given by (\ref{eqn:unblindedhazard}) while the dwindling blinded cohort would use hazard $h(t)$ given by 
(\ref{eqn:loglin_decay_haz}). 
Following the crossover interlude (\ref{eqn:unblindedhazard}) would apply 
to all and at some point, the term $\exp(\vect X \bbeta)$
might not be needed if the volunteers in the two arms behave similarly. This construction is a form of time-dependent stratification. 

A simpler way to address  open label crossover bias is to 
define a blackout period of length  $\Delta^*$  such that 
behavior is  presumed
to be similar after the end of the period. 
Similarity should happen
at some point as all trial volunteers will know they are vaccinated and protected.  As above, time dependent stratification would
make sense with different baseline hazards before unblinding and after the crossover blackout period.
To be specific, for the 
log-linear decay  model we would have 
$h(t) = \lambda_0(t) \exp\{ Z(t) [\theta_1 + \theta_2(t-\tau^{(v)})]\}$
prior to unblinding and 
$h^*(t) = \lambda_0^*(t)
\exp\{ \theta_2(t-\tau^{(v)})\}$ at time $\Delta^*$ post unblinding.  

If the VE dropped substantially during a black-out period, later estimates of VE might be compromised.
As an extreme example, suppose all volunteers enroll at the same time, and all are blacked out during  $\tau,\tau+\Delta^*$ which is exactly 
when VE drops. Then the estimated VE curve pre- and post-crossover would incorrectly appear constant. 
In practice this scenario can be avoided with a staggered entry trial by exploiting the  induced variation in time since vaccination at any calendar time.   To illustrate what not to do, suppose that  enrollment took 2 months, crossover took 2 months and the crossover order was in exact sync with the enrollment order.
Then all would be crossed over at some time $\tau$ since
randomization and all blacked out for the 
period $\tau,\tau+\Delta^*$. To minimize the problem, crossover could occur in reverse order with the first enrollees being crossed over last. Logistical considerations and placebo volunteers' sense of fairness
could also come into play.


\section{Assumptions}
To recover a VE curve under a standard trial with no crossover requires that the volunteers in each arm remain  similar over time and that the external environment remain   similar over time.   
\begin{itemize}
	\item  \textit{Volunteers in each arm remain similar over time.} This can be violated if there is differential dropout in the two arms and dropout is related to underlying risk of disease. Relatedly, unobserved heterogeneity in risk can result in differential culling by infection of the vaccine and placebo groups. Thus after a while, the remaining placebo arm volunteers tend to be a less risky group than the remaining vaccine arm  volunteers  and VE can appear to decrease, see \cite{lipsitch2019challenges,durham1998estimation,aalen2015does}. COVID-19 trials with 30,000 or more enrolled and perhaps 200-1000 cases over follow-up, such bias may be small. 
	Of course one can can  explicitly model the heterogeneity
	\citep{kanaan2002estimation}.  Such methods are beyond the scope of this paper.    
	In practice, covariate adjustment for baseline risk factors can be used to mitigate this assumption.   
	
	\item \textit{Study environment similar.} The proportional hazards model allows for the attack rate to change with time. But if the pathogen mutates to a form that is resistant to vaccine effects, efficacy may appear to wane. Another possibility is if human behavior changes in such a way that the vaccine is less effective. For example, 
	if there is less mask wearing in the community over the study, 
	the viral inoculum at infection may increase over the study and overwhelm the immune response for later cases. Vaccines may work less well against larger inoculums and thus VE might appear to wane.  For viral mutation, analyses could be run separately for different major strains provided they occur both prior and post crossover.  More elaborate methods could also be developed to address viral mutation, but are beyond the scope of this paper.          
\end{itemize}  

The only additional assumption that is required to recover the VE profile  under placebo crossover is that the effect of vaccination be the same no matter when the vaccine is given.  Interestingly, this is a common assumption; for vaccine trials with staggered entry it is implicitly assumed that  the VE for early enrollees  is the  same as for  the late enrollees. Importantly, both standard and crossover designs allow for a varying attack rate 
through the arbitrary hazard $h_0(t)$ whether due to seasonality, vaccination coverage, or other reasons.   

%

\section{Simulated COVID-19 Trials}
\label{sec:simulations}

In this section, we explore how placebo crossover, the dynamics of durability, and the baseline hazard affect our estimates of vaccine efficacy and durability. 
Since several COVID-19 vaccine trials are powered to accrue 150 cases and follow all
volunteers for two years, we evaluate three different designs:  i) crossover at 150 cases, ii)crossover at 1 year and iii) a 
standard parallel arm trial \citep{modernaprotocol}. 
We consider two settings for vaccine dynamics: constant VE of 75\%, and VE waning linearly on the log-hazard scale from 85\% to 35\% over 1.5 years. In the crossover scenarios, placebo arm volunteers cross over during a four week interlude. For each of the six settings we simulated 10,000 trials. Each trial enrolled 3,000 participants in a 1:1 randomization with linear accrual of participants over an initial three month period and followed participants for two years post enrollment.
While COVID-19 trials are larger, we evaluated 3,000 participants 
to lessen our computational burden.
The baseline hazard was piecewise-constant and calibrated to yield an average of 50, 75, 50, and 25 cases per three month period in the placebo arm in the first year, and either the same or half the year one case rates in the second study year. The data were analyzed in each simulation using the log-linear decay model, (\ref{eqn:loglin_decay_haz}), and the P-spline model, (\ref{eqn:pspline_haz}).  


The simulations demonstrate that we can accurately estimate VE(s) and the change in VE in all simulation settings using both the log-linear and P-spline model (Tables \ref{tab:sim_res_waninghaz} and \ref{tab:sim_res_samehaz}). Coverage probabilities of 95\% confidence intervals were near their nominal levels or somewhat conservative. The P-spline model performs similarly to the log-linear model except for the estimates at year two where the variance becomes notably larger. 
Initiating crossover at one year resulted in an average accrual of 44\% more cases prior to crossover compared with trials that initiated crossover at 150 cases. We found that this improved the precision of our estimates for all quantities of interest. One way to quantify the relative performance of placebo crossover and parallel arm trials is by the ratio of empirical variances. We focus on  the empirical variances of estimates of the linear predictor in the log-linear model in the constant VE(s) setting in Table \ref{tab:sim_res_waninghaz}.
The cross at 150:cross at one year variance ratios for $\widehat{\mbox{VE}}(s)$ are
0.051/0.029=1.8, 2.4, 2.5, and 2.4  at $0.5,\ 1,\ 1.5,\ $ and two years. This underscores the potential benefit of additional case accrual during the pre-crossover period leading into the second year when the baseline hazard was halved. We next compare the crossover at one year design to a standard parallel trial using the log-linear model. This comparison is more of a benchmark as a standard trial is may not be  ethically possible following vaccine approval. The analogous empirical variance ratios for cross at one year compared to a standard trial are 0.029/0.022=1.3, 2.5, 2.3 and 2.0 respectively.  
Results are broadly similar for the P-spline model and for waning VE.

%
%

%

\begin{table}[htbp]
	\centering\fontsize{9.5pt}{10pt}\selectfont
	\begin{tabular}{ccccccccc}
		\toprule
		
		\multicolumn{3}{c}{ } & \multicolumn{3}{c}{$\log(1 - VE(s))$} & \multicolumn{3}{c}{$\log(1 - VE(s)) - \log(1 -  VE(0))$} \\
		\cmidrule(l{3pt}r{3pt}){4-6} \cmidrule(l{3pt}r{3pt}){7-9}
		\textbf{Design} & \textbf{Model} & \textbf{Time} & \textbf{Bias} & \textbf{Emp. Var.} & \textbf{Covg.} & \textbf{Bias} & \textbf{Emp. Var.} & \textbf{Covg.}\\
		\midrule
		\addlinespace[0.3em]
		\multicolumn{9}{l}{\textbf{True vaccine efficacy constant at 75\%}}\\
		\cmidrule(lr){1-9}\hspace{1em}Cross at 150 cases & log-linear & 0.5 & -0.013 & 0.051 & 0.948 & 0.002 & 0.048 & 0.952\\
		\hspace{1em} $\tau_x = 0.6 \pm 0.05$ &  & 1.0 & -0.011 & 0.146 & 0.951 & 0.004 & 0.192 & 0.952\\
		\hspace{1em} &  & 1.5 & -0.009 & 0.338 & 0.953 & 0.006 & 0.433 & 0.952\\
		\hspace{1em} &  & 2.0 & -0.007 & 0.626 & 0.953 & 0.008 & 0.769 & 0.952\\
		\addlinespace[0.3em]\hspace{1em} & P-spline & 0.5 & -0.016 & 0.067 & 0.967 & 0.006 & 0.142 & 0.971\\
		\hspace{1em} &  & 1.0 & -0.017 & 0.194 & 0.965 & 0.004 & 0.288 & 0.959\\
		\hspace{1em} &  & 1.5 & -0.014 & 0.369 & 0.958 & 0.008 & 0.460 & 0.956\\
		\hspace{1em} &  & 2.0 & -0.015 & 0.951 & 0.974 & 0.006 & 1.043 & 0.972\\
		\cmidrule(lr){1-9}\hspace{1em}Cross at 1 year & log-linear & 0.5 & -0.009 & 0.029 & 0.949 & 0.001 & 0.022 & 0.953\\
		\hspace{1em} $N_x = 216 \pm 13$ &  & 1.0 & -0.009 & 0.061 & 0.953 & 0.001 & 0.087 & 0.953\\
		\hspace{1em} &  & 1.5 & -0.008 & 0.137 & 0.952 & 0.002 & 0.195 & 0.953\\
		\hspace{1em} &  & 2.0 & -0.007 & 0.256 & 0.953 & 0.002 & 0.347 & 0.953\\
		\addlinespace[0.3em]\hspace{1em} & P-spline & 0.5 & -0.013 & 0.041 & 0.982 & 0.005 & 0.146 & 0.978\\
		\hspace{1em} &  & 1.0 & -0.014 & 0.084 & 0.976 & 0.004 & 0.178 & 0.969\\
		\hspace{1em} &  & 1.5 & -0.010 & 0.160 & 0.969 & 0.008 & 0.240 & 0.972\\
		\hspace{1em} &  & 2.0 & -0.024 & 0.671 & 0.987 & -0.006 & 0.786 & 0.983\\
		\cmidrule(lr){1-9}\hspace{1em}Parallel trial & log-linear & 0.5 & -0.010 & 0.022 & 0.948 & -0.001 & 0.016 & 0.952\\
		\hspace{1em} &  & 1.0 & -0.010 & 0.024 & 0.951 & -0.001 & 0.065 & 0.952\\
		\hspace{1em} &  & 1.5 & -0.011 & 0.059 & 0.949 & -0.002 & 0.145 & 0.952\\
		\hspace{1em} &  & 2.0 & -0.011 & 0.125 & 0.949 & -0.002 & 0.258 & 0.952\\
		\addlinespace[0.3em]\hspace{1em} & P-spline & 0.5 & -0.012 & 0.039 & 0.981 & 0.014 & 0.173 & 0.979\\
		\hspace{1em} &  & 1.0 & -0.013 & 0.056 & 0.981 & 0.012 & 0.208 & 0.967\\
		\hspace{1em} &  & 1.5 & -0.024 & 0.073 & 0.980 & 0.002 & 0.194 & 0.974\\
		\hspace{1em} &  & 2.0 & -0.071 & 0.447 & 0.981 & -0.045 & 0.560 & 0.981\\
		\bottomrule\\
		\addlinespace[0.3em]
		\multicolumn{9}{l}{\textbf{Vaccine efficacy wanes from 85\% to 35\% over 1.5 years}}\\
		\cmidrule(lr){1-9}\hspace{1em}Cross at 150 cases & log-linear & 0.5 & -0.014 & 0.051 & 0.951 & 0.003 & 0.031 & 0.953\\
		\hspace{1em} $\tau_x = 0.63 \pm 0.05$ &  & 1.0 & -0.010 & 0.108 & 0.951 & 0.007 & 0.123 & 0.953\\
		\hspace{1em} &  & 1.5 & -0.007 & 0.225 & 0.951 & 0.010 & 0.276 & 0.953\\
		\hspace{1em} &  & 2.0 & -0.004 & 0.405 & 0.950 & 0.014 & 0.491 & 0.953\\
		\addlinespace[0.3em]\hspace{1em} & P-spline & 0.5 & -0.015 & 0.067 & 0.964 & 0.011 & 0.143 & 0.977\\
		\hspace{1em} &  & 1.0 & -0.008 & 0.162 & 0.965 & 0.018 & 0.277 & 0.964\\
		\hspace{1em} &  & 1.5 & 0.000 & 0.247 & 0.960 & 0.027 & 0.359 & 0.960\\
		\hspace{1em} &  & 2.0 & 0.005 & 0.475 & 0.962 & 0.031 & 0.587 & 0.962\\
		\cmidrule(lr){1-9}\hspace{1em}Cross at 1 year & log-linear & 0.5 & -0.010 & 0.031 & 0.950 & 0.004 & 0.017 & 0.949\\
		\hspace{1em} $N_x = 211 \pm 13$ &  & 1.0 & -0.006 & 0.053 & 0.952 & 0.008 & 0.066 & 0.949\\
		\hspace{1em} &  & 1.5 & -0.001 & 0.107 & 0.950 & 0.013 & 0.149 & 0.949\\
		\hspace{1em} &  & 2.0 & 0.003 & 0.195 & 0.951 & 0.017 & 0.265 & 0.949\\
		\addlinespace[0.3em]\hspace{1em} & P-spline & 0.5 & -0.013 & 0.039 & 0.976 & 0.014 & 0.142 & 0.983\\
		\hspace{1em} &  & 1.0 & -0.005 & 0.072 & 0.973 & 0.022 & 0.182 & 0.970\\
		\hspace{1em} &  & 1.5 & 0.008 & 0.119 & 0.966 & 0.035 & 0.208 & 0.973\\
		\hspace{1em} &  & 2.0 & 0.016 & 0.352 & 0.977 & 0.043 & 0.469 & 0.976\\
		\cmidrule(lr){1-9}\hspace{1em}Parallel trial & log-linear & 0.5 & -0.010 & 0.024 & 0.948 & 0.004 & 0.012 & 0.951\\
		\hspace{1em} &  & 1.0 & -0.006 & 0.016 & 0.946 & 0.008 & 0.048 & 0.951\\
		\hspace{1em} &  & 1.5 & -0.002 & 0.031 & 0.951 & 0.012 & 0.107 & 0.951\\
		\hspace{1em} &  & 2.0 & 0.002 & 0.071 & 0.950 & 0.015 & 0.191 & 0.951\\
		\addlinespace[0.3em]\hspace{1em} & P-spline & 0.5 & -0.012 & 0.036 & 0.982 & 0.022 & 0.164 & 0.984\\
		\hspace{1em} &  & 1.0 & -0.007 & 0.038 & 0.983 & 0.027 & 0.213 & 0.970\\
		\hspace{1em} &  & 1.5 & -0.004 & 0.039 & 0.980 & 0.030 & 0.182 & 0.975\\
		\hspace{1em} &  & 2.0 & -0.006 & 0.203 & 0.979 & 0.028 & 0.337 & 0.976\\
		\bottomrule\\\vspace{0.1em}
	\end{tabular}
	\caption{Bias, empirical variance, and coverage for the linear predictor in Cox PH models with time-varying VE in simulated trials where the baseline hazard in  year two was half the baseline hazard in  year one. The log-linear and P-spline models correspond to (\ref{eqn:loglin_decay_haz}) and (\ref{eqn:pspline_haz}), respectively. The average time of crossover (in years), $\tau_x$, and the average number of events at crossover, $N_x$, along with standard deviations beneath the crossover grouping in the design column. Time is given in years since study initiation.}
	\label{tab:sim_res_waninghaz}
\end{table}

In Tables \ref{tab:simres_par_ests}
and \ref{tab:simres_par_ests_samehaz} we
provide estimates of the intercept and linear trend of the VE profile for the scenarios where the baseline hazards
was the same or halved in year two, respectively.  
All estimates have negligible bias and good coverage.  
For the constant VE scenario and log-linear model, 
the variance ratios for cross at 150 cases and cross at one year are 0.051/0.040=1.3 and 0.192/0.084=2.3 for the intercept and  slope, respectively.   Thus there is  
a big advantage in slope estimation with delayed crossover.  When we compare crossover at one year to a standard trial, the  variance ratios are
.040/.052=0.8 and .087/.065=1.3 respectively. 
Interestingly, crossover improves the 
intercept estimate as during the crossover interlude, the newly vaccinated placebo volunteers contribute additional information about the intercept.  
The P-spline and log-linear model have similar empirical variances for the slope but the log-linear model has about half the empirical variance of the 
P-spline model for the  intercept.    
Finally, for the constant VE scenario
we compared the intercept estimates to a constant VE model (top half of Table \ref{tab:simres_par_ests}).  Under crossover, the empirical variance was modestly improved under this model compared to the log-linear model. Conclusions are broadly similar for  the waning VE scenario. 



\begin{table}[htbp]
	\centering \fontsize{9.5pt}{10pt}\selectfont
	\begin{tabular}{llcccccc}
		\toprule
		\multicolumn{2}{c}{ } & \multicolumn{3}{c}{\textbf{Intercept}} & \multicolumn{3}{c}{\textbf{Linear trend}} \\
		\cmidrule(l{3pt}r{3pt}){3-5} \cmidrule(l{3pt}r{3pt}){6-8}
		&  & \textbf{Bias} & \textbf{Emp. Var.} & \textbf{Covg.} & \textbf{Bias} & \textbf{Emp. Var.} & \textbf{Covg.}\\
		\midrule
		\addlinespace[0.3em]
		\multicolumn{6}{l}{\textbf{Vaccine efficacy constant at 75\%}}\\
		\addlinespace[0.3em]\hspace{1em}Cross at 150 cases & Constant VE & -0.012 & 0.039 & 0.952 & --- & --- & ---\\
		\addlinespace[0.3em]\hspace{1em}\hspace{1em} & log-linear & -0.015 & 0.051 & 0.951 & 0.004 & 0.192 & 0.952\\
		\hspace{1em} & P-spline & -0.022 & 0.086 & 0.973 & 0.005 & 0.188 & 0.956\\
		\addlinespace[0.3em]\hspace{1em}Cross at 1 year & Constant VE & -0.008 & 0.028 & 0.950 & --- & --- & ---\\
		\hspace{1em} & log-linear & -0.010 & 0.040 & 0.950 & 0.001 & 0.087 & 0.953\\
		\hspace{1em} & P-spline & -0.018 & 0.104 & 0.978 & 0.001 & 0.084 & 0.959\\
		\addlinespace[0.3em]\hspace{1em}Parallel trial & Constant VE & -0.005 & 0.018 & 0.949 & --- & --- & ---\\
		\hspace{1em} & log-linear & -0.009 & 0.052 & 0.951 & -0.001 & 0.065 & 0.952\\
		\hspace{1em} & P-spline & -0.026 & 0.126 & 0.974 & 0.006 & 0.063 & 0.955\\
		\bottomrule\\
		\addlinespace[0.3em]
		\multicolumn{8}{l}{\textbf{Vaccine efficacy wanes from 85\% to 35\% over 1.5 years}}\\
		\addlinespace[0.3em]\hspace{1em}Cross at 150 cases & log-linear & -0.017 & 0.056 & 0.951 & 0.007 & 0.123 & 0.953\\
		\hspace{1em} & P-spline & -0.027 & 0.102 & 0.975 & -0.002 & 0.121 & 0.955\\
		\addlinespace[0.3em]\hspace{1em}Cross at 1 year & log-linear & -0.014 & 0.043 & 0.952 & 0.008 & 0.066 & 0.949\\
		\hspace{1em} & P-spline & -0.027 & 0.116 & 0.981 & -0.003 & 0.065 & 0.951\\
		\addlinespace[0.3em]\hspace{1em}Parallel trial & log-linear & -0.014 & 0.057 & 0.950 & 0.008 & 0.048 & 0.951\\
		\hspace{1em} & P-spline & -0.034 & 0.145 & 0.978 & 0.003 & 0.047 & 0.952\\
		\bottomrule\\\vspace{0.1em}
	\end{tabular} 
	\caption{Empirical variance and coverage for estimates of the intercept and linear trend in vaccine efficacy under the log--linear model, (\ref{eqn:loglin_decay_haz}), and semi--parametric model, (\ref{eqn:pspline_haz}). Here, the time--varying baseline hazard in year two was half the baseline hazard in  year one.}
	\label{tab:simres_par_ests}
\end{table}

\subsection{Crossover Interlude}
A design question is how estimation efficiency varies with the length of the crossover interlude.  To explore this design question, we did additional simulations where we evaluated a standard parallel trial of two years, a trial where all placebo participants are crossed over at one year, and a trial where the times of vaccination for all volunteers were uniformly distributed over two years. 
The baseline hazard was constant over the two year period. Under the constant VE(s) scenario, the empirical variances for the intercept term were  0.051, 0.035, 0.031, respectively while  the variances for the slope were 0.039, 0.039, and 0.034 respectively (Table \ref{tab:sim_res_idealized}).
This suggests a longer crossover interlude is somewhat better for estimation of the intercept and the slope.

\subsection{Unobserved Heterogeneity in Risk} 
Unobserved heterogeneity in the risk of disease can lead to bias in estimates of VE(s) and complicate the task of separating time-varying efficacy from 
increased removal of the riskier individuals from the 
placebo arm \citep{balan2020tutorial}. We simulated placebo crossover and parallel arm trials with 30,000 participants and gamma distributed frailties with mean one, and variance equal to either one or four. Crossover trials initiated vaccination of the placebo arm at one year. The baseline hazard was constant and calibrated to yield either 50 or 300 cases per six month period on the placebo arm, and VE(s) was either constant or waned linearly on the log hazard scale, as before. The frailty distributions in the original placebo and vaccine arms at the end of followup were more similar in the placebo crossover trials than in the standard parallel trials (Table \ref{tab:frailty_sumstats}). In the low baseline hazard scenario, where the dominant contribution to a participant's propensity for disease was their underlying frailty, placebo crossover trials yielded less biased estimates of VE(s) relative to the standard parallel design (Tables \ref{tab:ve_frailty_const} and \ref{tab:ve_frailty_waning}). 
Higher baseline hazards resulted in more differential culling of the risk set and increased bias in estimates of VE(s).
In the high baseline hazard scenario, the common baseline hazard dominated heterogeneity in the frailty distribution, and in this setting the bias in VE(s) estimates under placebo crossover was comparable to the bias that was observed with parallel trials.
However, the absolute bias was modest.  Under a constant
VE of 75\%, the estimates at two years were about 70\% on average under either design.
This scenario with a variance of four had  substantial heterogeneity in the risk of disease with the riskiest 1\% in the placebo arm having a median probability of disease over two years of 0.60.
The least risky 1\% had a median probability of $1 \times 10^{-10}$. In practice, we could mitigate biases resulting from heterogeneity in the frailty distribution by adjusting for known risk factors of disease and stratifying our analyses by site or geographic region. 

\section{Analysis of Two Simulated Trials}
\label{sec:example}
In this section, we present detailed analyses of two simulated COVID-19 vaccine trials where the true VE profile was either constant at 75\% or waned linearly on the log--hazard scale from 85\% to 35\% over 1.5 years. Each trial enrolled 30,000 participants with linear accrual over three months in a 1:1 randomization to vaccine or placebo. The baseline attack rate was piecewise constant with changepoints every three months, and was calibrated to yield 50, 75, 50, and 25 cases on the placebo arm in each period in the first year, and half the expected number of cases per period on the placebo arm in year two. In this example, interim analyses are planned at 150 cases, which ultimately result in crossover at the end of year one following evaluation and vetting of the efficacy by a regulatory agency. Placebo crossover occurs over a four week period. Each volunteer was followed for a total of two years. 

The two simulated trials are summarized in Table \ref{tab:example_simstats}. In the constant VE scenario, the trial reached 150 cases in 222 days, and recorded 223 events by the one year crossover time--point and 273 events, overall. In the waning VE scenario, the trial reached 150 cases in 242 days, and recorded 199 events by the one year crossover time--point and 292 events, overall. The case split across treatment arms declined from roughly 83\% on the placebo arm at the 150 case interim look to 76.2\% at the completion of the study in the constant VE scenario, and from 82\% to 65.2\% in the waning VE scenario. The overall VE estimate at the one year crossover, estimated using a proportional hazards model without adjustment for time since vaccination, was 76.6\% (95\% CI: 67.2\%, 83.3\%) in the constant VE(s) case and 80.1\% (95\% CI: 71.0\%, 86.3\%) in the waning VE(s) scenario (the true geometric mean VE(s) to one year post--vaccination is 75.6\%). 

\begin{figure}
	\centering
	\includegraphics[width=\linewidth]{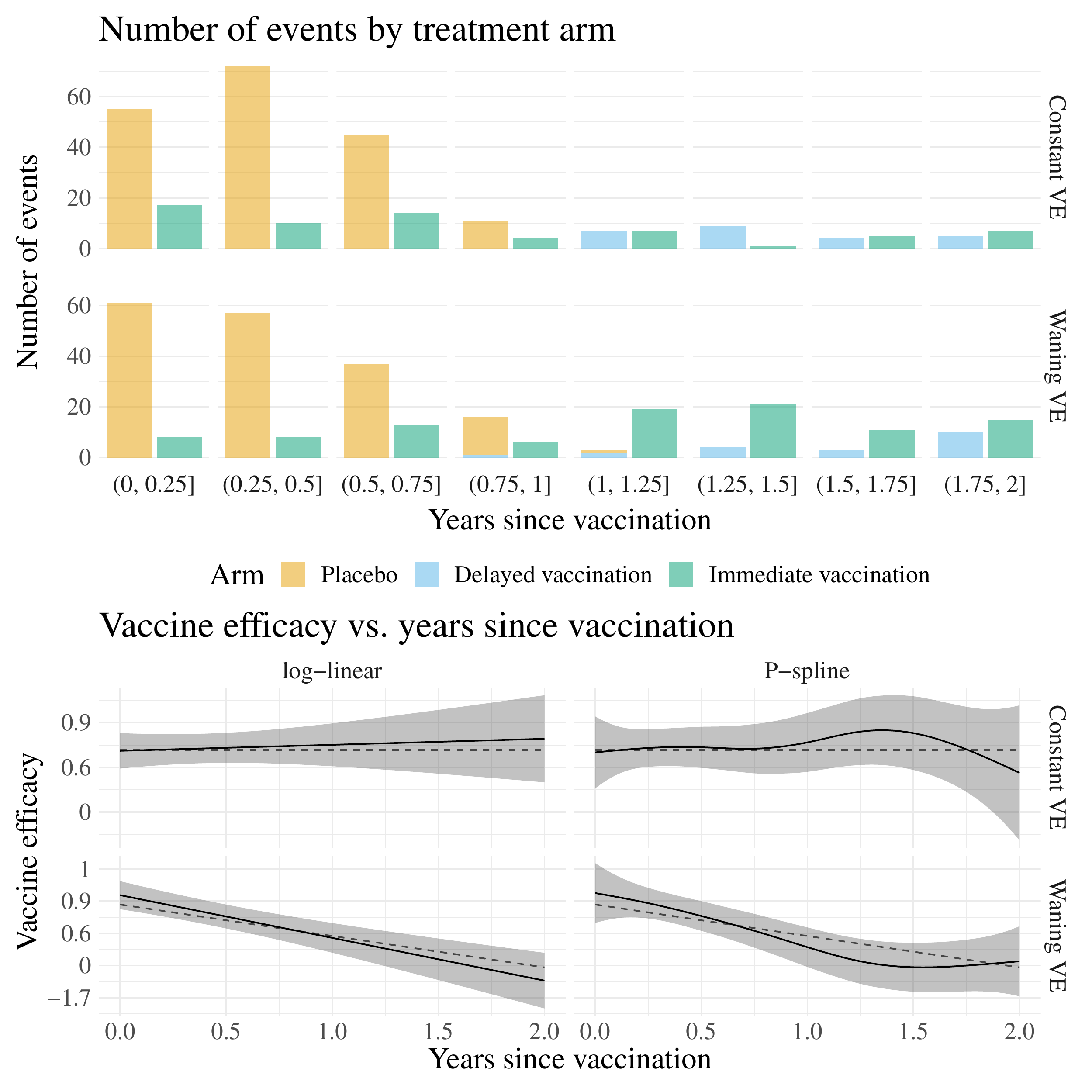}
	\caption{(Top) Number of events per quarter by arm. The delayed vaccination arm consists of the original placebo participants after they have been crossed over. (Bottom) Vaccine efficacy (VE) as a function of time since vaccination. Dashed lines are the true VE(s), solid curves and ribbons are pointwise means and 95\% confidence intervals.}
	\label{fig:VE_plot}
\end{figure}

Point estimates for VE(0) and the linear trend in log VE(s) from the log--linear and P-spline models were close in both scenarios, although confidence intervals in the P-spline models were wider. The estimated efficacy profiles obtained with both methods were in agreement and recovered the true VE profile (Figure \ref{fig:VE_plot}). The P-spline estimates had wider point-wise confidence intervals, but the inflation in the variance appears to be fairly modest for the period spanning the end of study enrollment through, roughly, year 1.5 post--vaccination. In practice, both the log--linear decay model and the P-spline model could be used to test a hypothesis of time--varying VE(s). This is straightforwardly carried out for the log--linear model via a likelihood ratio test (LRT) for the slope parameter in (\ref{eqn:loglin_decay_haz}) where the test statistic is compared to a chi--square distribution with one degree of freedom. For the P-spline models, we perform a likelihood ratio test for whether all of the P-spline basis coefficients are jointly equal to zero, and compare the test statistic to a chi--square distribution with 3.1 degrees of freedom (the effective degrees of freedom for the P-splines in our models). In the waning VE(s) scenario, we resoundingly 
reject the null hypothesis of time-homogeneous VE(s), and fail to reject the null in the constant VE(s) scenario (Table \ref{tab:example_simstats}) at the end of two years of follow-up.
The benefit of an additional year of follow-up past crossover is substantial
in terms of evaluating the long term durability of the vaccine.  Under the waning 
VE scenario, the p-value for testing the null hypothesis of constant VE is
close to 0.05 at one year and convincing at two years for both the log-linear and P-spline models.     
These simulated examples show that for both the waning and constant VE scenarios, accurate inference about the behavior of the VE over time can be recovered. 

\begin{table}[htbp]
	\centering\fontsize{9.5pt}{10pt}\selectfont
	\begin{tabular}{lcc}
		\toprule
		& \textbf{True VE Constant at 75\%} & \textbf{True VE Wanes from 85\% to 35\%}  \\
		\cmidrule(lr){2-2} \cmidrule(lr){3-3}
		\textbf{Time of 150 case interim look} & Day 222 & Day 242 \\
		\addlinespace[0.3em]
		\textbf{Case split by original arm} && \\
		\hspace{1em} \textit{at interim look} & Placebo = 124, Vaccine = 26 & Placebo = 131, Vaccine = 19 \\
		\hspace{1em} \textit{at 1 year crossover} & Placebo = 181, Vaccine = 42 & Placebo = 166, Vaccine = 33 \\
		\hspace{1em} \textit{at 2 year follow-up} & Placebo = 208, Vaccine = 65 & Placebo = 191, Vaccine = 101\\
		\addlinespace[0.3em]
		\textbf{Estimates at interim look} & & \\
		\hspace{1em}\textit{log-linear model} & & \\
		\hspace{1em}\hspace{1em}Intercept & -0.84 (95\% CI: -1.6, -0.09) & -2.16 (95\% CI: -3.17, -1.16)\\
		\hspace{1em}\hspace{1em}Linear trend & -3.06 (95\% CI: -6.05, -0.07) & 0.81 (95\% CI: -2.13, 3.75)\\
		\hspace{1em}\hspace{1em}LRT for time-varying VE & 0.039 & 0.589\\
		\addlinespace[0.3em]
		\hspace{1em}\textit{P-spline model} & &\\
		\hspace{1em}\hspace{1em}Intercept & -1.41 (95\% CI: -2.77, -0.05) & -2.43 (95\% CI: -4.23, -0.62)\\
		\hspace{1em}\hspace{1em}Linear trend & -3.02 (95\% CI: -6.36, 0.32) & 0.8 (95\% CI: -2.13, 3.73)\\
		\hspace{1em}\hspace{1em}LRT for time-varying VE & 0.037 & 0.605\\
		\addlinespace[0.3em]
		\multicolumn{3}{l}{\textbf{Estimates at 1 year crossover}}\\
		\hspace{1em}\textit{log-linear model} & & \\
		\hspace{1em}\hspace{1em}Intercept & -1.34 (95\% CI: -1.98, -0.7) & -2.36 (95\% CI: -3.17, -1.55)\\
		\hspace{1em}\hspace{1em}Linear trend & -0.29 (95\% CI: -1.74, 1.17) & 1.8 (95\% CI: 0.2, 3.4)\\
		\hspace{1em}\hspace{1em}LRT for time-varying VE & 0.698 & 0.027\\
		\addlinespace[0.3em]
		\hspace{1em}\textit{P-spline model} & &\\
		\hspace{1em}\hspace{1em}Intercept & -1.14 (95\% CI: -2.17, -0.1) & -2.26 (95\% CI: -3.68, -0.83)\\
		\hspace{1em}\hspace{1em}Linear trend & -0.28 (95\% CI: -1.66, 1.1) & 1.8 (95\% CI: 0.22, 3.37)\\
		\hspace{1em}\hspace{1em}LRT for time-varying VE & 0.133 & 0.054\\
		\addlinespace[0.3em]
		\multicolumn{3}{l}{\textbf{Estimates at 2 year follow-up}}\\
		\hspace{1em}\textit{log-linear model} & & \\
		\hspace{1em}\hspace{1em}Intercept & -1.37 (95\% CI: -1.77, -0.97) & -2.19 (95\% CI: -2.62, -1.75)\\
		\hspace{1em}\hspace{1em}Linear trend & -0.13 (95\% CI: -0.7, 0.43) & 1.33 (95\% CI: 0.82, 1.83)\\
		\hspace{1em}\hspace{1em}LRT for time-varying VE & 0.641 & <0.001\\
		\addlinespace[0.3em]
		\hspace{1em}\textit{P-spline model} & &\\
		\hspace{1em}\hspace{1em}Intercept & -1.33 (95\% CI: -2.09, -0.58) & -2.26 (95\% CI: -3.15, -1.36)\\
		\hspace{1em}\hspace{1em}Linear trend & -0.13 (95\% CI: -0.7, 0.44) & 1.28 (95\% CI: 0.77, 1.8)\\
		\hspace{1em}\hspace{1em}LRT for time-varying VE & 0.178 & <0.001\\
		\bottomrule\\\vspace{0.1em}
	\end{tabular}
	\caption{Summary of example trials simulated under constant and waning vaccine efficacy (VE) at times of interim analysis and placebo crossover. The intercept and linear trend  correspond to the immediate effect of vaccination and the time--trend for VE(s) under model (\ref{eqn:loglin_decay_haz}), and the true values were set to $\theta_1 = -1.39$ and $\theta_2 = 0$ in the constant VE scenario, and $\theta_1 = -1.9$ and $\theta_2 = 0.98$ in the waning VE setting. The likelihood ratio test (LRT) for waning VE compares models (\ref{eqn:loglin_decay_haz}) and (\ref{eqn:pspline_haz}) to a PH model without adjustment for time since vaccination.}
	\label{tab:example_simstats}
\end{table}

\section{Discussion}

Knowing the durability of vaccine induced protection is a key question in vaccine development, especially for COVID-19 vaccines. With  placebo volunteers being offered  vaccine before long term follow-up has completed, it seems the ability to assess durability is lost.  In this paper we demonstrated that placebo controlled VE can be accurately assessed long after the placebo group has disappeared. Our method is the familiar Cox proportional hazards model. To reflect seasonal or outbreak variation in the attack rate, we used calendar time as the time index.   To recover different VE curves we  specified flexible models for VE decay. If crossover occurs quickly, the early VE(s) will remain poorly estimated, no matter how many post-crossover cases occur which will impact later estimates of VE(s). Our results point out the advantages of delaying crossover and longer crossover interludes to help improve the estimation. We also provide suggestions on how to manage the crossover interlude, discuss how to perform per-protocol analyses, and discuss solutions for open label crossover where risk behavior might increase for the recently unblinded vaccinees. While developing this approach we became aware of two related methods that assess vaccine durability. Both use hazard function models with calendar time as the index and consider the pre-crossover \cite{lin2021evaluating}
and both the pre and postcrossover period \cite{tsiatis2021estimating}.  The latter focuses on
methods when crossover is open label and confounding an issue.


Future work could develop random effects or frailty type models. Such models seem especially suited for  settings with a relatively high attack rate.  Our work focused on the setting where the disease event was continuously monitored.  An important endpoint in vaccine trials is seroconversion, or the development of antibodies to 
the pathogen of interest. Seroconversion is typically measured rarely which results in an interval censored endpoint. The extension of these methods to interval censored data will be important.
Another potential generalization of these methods is for observational data where
methods could be developed that address confounding of risk with vaccine uptake.  
Finally, an emerging issue in the context of COVID-19 is how to estimate vaccine durability in the presence of emerging strains. This could be addressed within a competing risks framework in which times to first acquiring disease due to different strains are treated as competing events. The framework developed in this paper are easily extended to this setting since our models could straightforwardly be applied to the sub-distribution hazards in a competing risks model.

\section*{Acknowledgments}
This work utilized the computational resources of the NIH HPC Biowulf computing cluster (http://hpc.nih.gov). The authors would like to thank Keith Lumbard for help with simulations, as well as Michael Fay, Anastasios Tsiatis, Danyu Lin, Peter Gilbert, Holly Janes,  and Larry Molton for helpful discussions regarding this work.

\section*{Supplementary Materials}
Code demonstrating how to simulate data and reproduce the results presented in this manuscript is made available in the following GitHub repository: \url{https://github.com/fintzij/ve_placebo_crossover}. A minimal working example with \texttt{R} and \texttt{SAS} code is provided in the appendix.

\vspace*{-8pt}
\bibliographystyle{plainnat}
\bibliography{deanbib}

\newpage

\appendix

\section{Illustrative Computer Code}

In this section we present a minimal example with \texttt{SAS} and \texttt{R} code to 
estimate a log-linear waning efficacy curve.
In this trial, a per-protocol analysis is used and disease 
cases are counted starting 30 days after the first dose.
Calendar time is relative to 1 January 2021,
so the volunteer depicted in the first row was dosed on 5 January 2021.  
Thus during the crossover period, cases are not counted for 30 days.
A blinded crossover
is assumed so the same placebo baseline hazard applies throughout the
study without time-dependent stratification.   If an open label
crossover were pursued, an
additional 'stratum' variable could be created that
identified whether a risk interval was blinded or open label.
The 'stratum' variable would then be used as a stratification variable
in the proportional hazards model.

We assume the data is arranged so that anyone who gets the crossover
dose has both the start and end date of crossover recorded and that if
someone drops out or has an event before the start of the crossover
period the start and end date of crossover are missing.

Volunteers 1, 2, 6, and 7 are censored after crossover while
Volunteer 4 has an event after crossover.
Volunteer 3 is censored before crossover while volunteers  5 and  8
have events before crossover. Volunteers 9 and 10 enter the crossover
interlude but have an event and dropout, respectively, thus they are
censored at the start of the crossover interlude.

The variables below are
\begin{verbatim}
	id         = subject identfier
	arm        = original randomization arm  1=vaccine 0=placebo
	entry      = # of days from 01-Jan-2021 to 30 days past 1st dose
	Xstart     = # of days from 01-Jan-2021 to crossover start
	Xend       = # of days from 01-Jan-2021 to 30 days post 1st crossover dose  
	eveenttime = # of days from 01-Jan-2021 to a disease event or censoring
	status     = 1 if a disease event at eventtime 0 otherwise 
\end{verbatim}

\subsection{SAS code}
\single
{
	\begin{verbatim}
		
		
		DATA new;
		INPUT  id	arm	entry	Xstart	Xend eventtime 	status;
		CARDS;
		1	0	35	65   95	370	0
		2	1	45	80  110	400	0
		3	0	55	 .   .  150	0
		4	1	60	170	200 310	1
		5	0	65	 .   .   80	1
		6	1	80	190	210 410	0
		7	0	85	215	245 420	0
		8	1	70   .   .  90 1
		9	0	58	160 190 180 1
		10	1	71	160 190 166 0
		;
		
		/* did not start crossover*/
		DATA data1; SET new;
		IF Xstart=. AND Xend = .;
		period=1;  start=entry; stop=eventtime; event=status;
		
		/* event or dropout during crossover interlude: censor at start of period 1*/
		DATA data2; SET new;
		IF Xstart^=. AND Xstart<=eventtime AND eventtime<=Xend;
		period=1; start=entry; stop=Xstart; event=0;
		
		/* did pass crossover so outputs for per and post-crossover periods*/
		DATA data3; SET new;
		IF Xstart^=. AND eventtime>Xend;
		period=1; start=entry; stop=Xstart;    event=0; OUTPUT;
		period=2; start=Xend;  stop=eventtime; event=status; OUTPUT;
		
		
		/* Merge the three datasets and mark the vaccination time and status*/
		DATA newest;
		SET data1 data2 data3;
		IF arm=0 THEN DO; timevact=Xend;  IF period=1 THEN vac=0; ELSE vac=1; END;
		IF arm=1 THEN DO; timevact=entry;                              vac=1; END;
		
		/* Run the code with a log-linear VE decay, fix so no missing vactime*/
		PROC PHREG DATA=newest;
		MODEL (start, stop)*event( 0 )=  vac  vactime
		/ itprint rl ;
		vactime=vac*(stop-timevact);
		IF vac=0 THEN vactime=0;
		RUN;
		
		/* Print out the dataset, sorted*/
		PROC SORT DATA=newest;
		BY id period;
		PROC PRINT;
		RUN;
		
		
		Obs id arm period start stop event vac timevact 
		1  1   0     1      35    65  0    0    95 
		2  1   0     2      95   370  0    1    95 
		3  2   1     1      45    80  0    1    45 
		4  2   1     2     110   400  0    1    45 
		5  3   0     1      55   150  0    0    . 
		6  4   1     1      60   170  0    1    60 
		7  4   1     2     200   310  1    1    60 
		8  5   0     1      65    80  1    0    . 
		9  6   1     1      80   190  0    1    80 
		10  6   1     2     210   410  0    1    80 
		11  7   0     1      85   215  0    0   245 
		12  7   0     2     245   420  0    1   245 
		13  8   1     1      70    90  1    1    70 
		14  9   0     1      58   160  0    0   190 
		15 10   1     1      71   160  0    1    71 
		
		
\end{verbatim}}
In this minimal example, the estimates of $(\theta_1,\theta_2)$ are $(-0.82335,0.02649)$,
respectively, which correspond to estimates of $VE(0),VE(30)$ of
$1-\exp(-0.82335)=0.561$ and $1-\exp(-0.82335+30 \times 0.02649)=0.028$. 

\subsection{R code}

{
	\begin{verbatim}
		library(survival)
		library(tidyverse) # for dplyr and piping %>%
		dean_dat = read.table("manuscript/dean_dat.txt")
		
		# dataset
		dat_raw = 
		data.frame(
		id        = c(1,2,3,4,5,6,7,8,9,10),
		arm       = c(0,1,0,1,0,1,0,1,0,1),        
		entry     = c(35,45,55,60,65,80,85,70,58,71), 
		Xstart    = c(65,80,NA,170,NA,190,215,NA,160,160),
		Xend      = c(95,110,NA,200,NA,210,245,NA,190,190),
		eventtime = c(370,400,150,310,80,410,420,90,180,166),
		status    = c(0,0,0,1,1,0,0,1,1,0)
		)
		
		# reshape dataset into start-stop format
		# add new variables for time-varying vaccine status and vaccination time
		dat_long = 
		dat_raw %>% 
		group_by(id) %>%
		summarize(id = rep(id, 2),
		arm = rep(arm, 2),
		tstart = c(entry, Xend),
		tstop = c(min(eventtime, Xstart, na.rm = T), eventtime),
		status = 
		case_when(is.na(Xstart) ~ 
		c(status, NA), # event happens before crossover
		between(eventtime, Xstart, Xend) ~ 
		c(0, NA),      # no event, then blackout period
		eventtime > Xend ~ 
		c(0, status)), # no event, then record status
		vacc_status = 
		case_when(arm == 1 ~ rep(1, 2),
		arm == 0 ~ c(0, 1)),
		vacc_time = 
		case_when(arm == 1 ~ rep(entry, 2),
		arm == 0 & is.na(Xstart) & is.na(Xend) ~ Inf,
		arm == 0 & !is.na(Xstart) & !is.na(Xend) ~ rep(Xend, 2))) %>% 
		drop_na() %>% 
		as.data.frame()
		
		print(dat_long, row.names = FALSE)
		# id  arm tstart tstop status vacc_status vacc_time
		# 1    0     35    65      0           0        95
		# 1    0     95   370      0           1        95
		# 2    1     45    80      0           1        45
		# 2    1    110   400      0           1        45
		# 3    0     55   150      0           0       Inf
		# 4    1     60   170      0           1        60
		# 4    1    200   310      1           1        60
		# 5    0     65    80      1           0       Inf
		# 6    1     80   190      0           1        80
		# 6    1    210   410      0           1        80
		# 7    0     85   215      0           0       245
		# 7    0    245   420      0           1       245
		# 8    1     70    90      1           1        70
		# 9    0     58   160      0           0       190
		# 10   1     71   160      0           1        71
		
		# fit that model!
		vacc_dur = 
		coxph(formula = 
		Surv(time  = tstart, 
		time2 = tstop,
		event =  status) ~ vacc_status + tt(vacc_time),
		tt = function(vacc_time, t, ...) {
			pmax(0, t - vacc_time)
		},
		data = dat_long)
\end{verbatim}}

The parameter estimates are $(\hat{\theta}_1,\ \hat{\theta_2})$ = (-0.82336, 0.02649).

\newpage
\section{Additional Simulation Results}

\subsection{Trials with Year Two Baseline Hazard Equal to Year One Baseline Hazard}
In this section we provide simulation results where the baseline
hazard function in year two is the same as in year one. 
Table \ref{tab:sim_res_samehaz} is the analogue to 
Table  1 and  Table
\ref{tab:simres_par_ests_samehaz} is the analogue to 
Table 2.

\begin{table}[htbp]
	\centering\fontsize{9.5pt}{10pt}\selectfont
	\begin{tabular}{ccccccccc}
		\toprule
		\multicolumn{3}{c}{ } & \multicolumn{3}{c}{$\log(1 - VE(s))$} & \multicolumn{3}{c}{$\log(1 - VE(s)) - \log(1 - VE(0))$} \\
		\cmidrule(l{3pt}r{3pt}){4-6} \cmidrule(l{3pt}r{3pt}){7-9}
		\textbf{Design} & \textbf{Model} & \textbf{Time} & \textbf{Bias} & \textbf{Emp. Var.} & \textbf{Covg.} & \textbf{Bias} & \textbf{Emp. Var.} & \textbf{Covg.}\\
		\midrule
		\addlinespace[0.3em]
		\multicolumn{9}{l}{\textbf{True vaccine efficacy constant at 75\%}}\\
		\cmidrule(lr){1-9}\hspace{1em}Cross at 150 cases & log-linear & 0.5 & -0.012 & 0.046 & 0.950 & 0.001 & 0.028 & 0.952\\
		\hspace{1em} $\tau_x = 0.6 \pm 0.05$ &  & 1.0 & -0.011 & 0.102 & 0.951 & 0.002 & 0.111 & 0.952\\
		\hspace{1em} &  & 1.5 & -0.010 & 0.213 & 0.950 & 0.003 & 0.249 & 0.952\\
		\hspace{1em} &  & 2.0 & -0.010 & 0.379 & 0.951 & 0.003 & 0.443 & 0.952\\
		\hspace{1em} & P-spline & 0.5 & -0.015 & 0.064 & 0.962 & 0.003 & 0.118 & 0.975\\
		\hspace{1em} &  & 1.0 & -0.014 & 0.148 & 0.960 & 0.004 & 0.229 & 0.960\\
		\hspace{1em} &  & 1.5 & -0.013 & 0.226 & 0.957 & 0.005 & 0.303 & 0.957\\
		\hspace{1em} &  & 2.0 & -0.019 & 0.461 & 0.966 & -0.001 & 0.536 & 0.963\\
		\cmidrule(lr){1-9}\hspace{1em}Cross at 1 year & log-linear & 0.5 & -0.008 & 0.027 & 0.950 & -0.001 & 0.011 & 0.953\\
		\hspace{1em} $N_x = 216 \pm 13$ &  & 1.0 & -0.009 & 0.043 & 0.951 & -0.002 & 0.042 & 0.953\\
		\hspace{1em} &  & 1.5 & -0.010 & 0.080 & 0.953 & -0.002 & 0.095 & 0.953\\
		\hspace{1em} &  & 2.0 & -0.011 & 0.138 & 0.953 & -0.003 & 0.170 & 0.953\\
		\hspace{1em} & P-spline & 0.5 & -0.010 & 0.035 & 0.976 & 0.005 & 0.101 & 0.981\\
		\hspace{1em} &  & 1.0 & -0.010 & 0.056 & 0.971 & 0.005 & 0.117 & 0.965\\
		\hspace{1em} &  & 1.5 & -0.013 & 0.087 & 0.969 & 0.002 & 0.132 & 0.977\\
		\hspace{1em} &  & 2.0 & -0.021 & 0.271 & 0.977 & -0.005 & 0.336 & 0.977\\
		\cmidrule(lr){1-9}\hspace{1em}Parallel trial & log-linear & 0.5 & -0.008 & 0.020 & 0.948 & 0.000 & 0.010 & 0.950\\
		\hspace{1em} &  & 1.0 & -0.008 & 0.013 & 0.950 & 0.000 & 0.040 & 0.950\\
		\hspace{1em} &  & 1.5 & -0.008 & 0.027 & 0.949 & 0.000 & 0.089 & 0.950\\
		\hspace{1em} &  & 2.0 & -0.008 & 0.060 & 0.949 & 0.000 & 0.158 & 0.950\\
		\hspace{1em} & P-spline & 0.5 & -0.010 & 0.033 & 0.981 & 0.012 & 0.134 & 0.983\\
		\hspace{1em} &  & 1.0 & -0.007 & 0.033 & 0.983 & 0.016 & 0.164 & 0.968\\
		\hspace{1em} &  & 1.5 & -0.012 & 0.035 & 0.981 & 0.011 & 0.142 & 0.975\\
		\hspace{1em} &  & 2.0 & -0.039 & 0.171 & 0.977 & -0.016 & 0.274 & 0.976\\
		\bottomrule\\
		\addlinespace[0.3em]
		\multicolumn{9}{l}{\textbf{Vaccine efficacy wanes from 85\% to 35\% over 1.5 years}}\\
		\cmidrule(lr){1-9}\hspace{1em}Cross at 150 cases & log-linear & 0.5 & -0.015 & 0.048 & 0.952 & 0.001 & 0.016 & 0.952\\
		\hspace{1em} $\tau_x = 0.63 \pm 0.05$ &  & 1.0 & -0.014 & 0.078 & 0.949 & 0.002 & 0.064 & 0.952\\
		\hspace{1em} &  & 1.5 & -0.013 & 0.140 & 0.950 & 0.004 & 0.144 & 0.952\\
		\hspace{1em} &  & 2.0 & -0.012 & 0.234 & 0.949 & 0.005 & 0.255 & 0.952\\
		\hspace{1em} & P-spline & 0.5 & -0.015 & 0.063 & 0.963 & 0.009 & 0.108 & 0.981\\
		\hspace{1em} &  & 1.0 & -0.010 & 0.127 & 0.962 & 0.015 & 0.208 & 0.968\\
		\hspace{1em} &  & 1.5 & -0.008 & 0.167 & 0.953 & 0.017 & 0.243 & 0.965\\
		\hspace{1em} &  & 2.0 & -0.007 & 0.254 & 0.959 & 0.018 & 0.329 & 0.962\\
		\cmidrule(lr){1-9}\hspace{1em}Cross at 1 year & log-linear & 0.5 & -0.009 & 0.030 & 0.948 & 0.001 & 0.008 & 0.953\\
		\hspace{1em} $N_x = 211 \pm 13$ &  & 1.0 & -0.008 & 0.039 & 0.952 & 0.003 & 0.031 & 0.953\\
		\hspace{1em} &  & 1.5 & -0.006 & 0.065 & 0.952 & 0.004 & 0.071 & 0.953\\
		\hspace{1em} &  & 2.0 & -0.005 & 0.106 & 0.954 & 0.006 & 0.126 & 0.953\\
		\hspace{1em} & P-spline & 0.5 & -0.010 & 0.034 & 0.972 & 0.012 & 0.089 & 0.987\\
		\hspace{1em} &  & 1.0 & -0.006 & 0.051 & 0.968 & 0.016 & 0.116 & 0.969\\
		\hspace{1em} &  & 1.5 & -0.002 & 0.070 & 0.966 & 0.020 & 0.118 & 0.980\\
		\hspace{1em} &  & 2.0 & 0.001 & 0.154 & 0.973 & 0.023 & 0.218 & 0.974\\
		\cmidrule(lr){1-9}\hspace{1em}Parallel trial & log-linear & 0.5 & -0.008 & 0.022 & 0.952 & 0.004 & 0.008 & 0.948\\
		\hspace{1em} &  & 1.0 & -0.004 & 0.010 & 0.947 & 0.007 & 0.031 & 0.948\\
		\hspace{1em} &  & 1.5 & -0.001 & 0.014 & 0.947 & 0.011 & 0.071 & 0.948\\
		\hspace{1em} &  & 2.0 & 0.003 & 0.034 & 0.949 & 0.014 & 0.125 & 0.948\\
		\hspace{1em} & P-spline & 0.5 & -0.011 & 0.030 & 0.980 & 0.019 & 0.119 & 0.990\\
		\hspace{1em} &  & 1.0 & -0.004 & 0.023 & 0.983 & 0.026 & 0.166 & 0.973\\
		\hspace{1em} &  & 1.5 & -0.001 & 0.020 & 0.978 & 0.029 & 0.144 & 0.976\\
		\hspace{1em} &  & 2.0 & -0.003 & 0.076 & 0.975 & 0.027 & 0.197 & 0.977\\
		\bottomrule\\\vspace{0.1em}
	\end{tabular}
	\caption{Bias, empirical variance, and coverage for the linear predictor in Cox PH models for simulated trials where the baseline hazard in  year two was the same as the baseline hazard in  year one. The log-linear and P-spline models correspond to (\ref{eqn:loglin_decay_haz}) and (\ref{eqn:pspline_haz}), respectively. The average time of crossover (in years), $\tau_x$, and the average number of events at crossover, $N_x$, along with standard deviations beneath the crossover grouping in the design column. Time is given in years since study initiation.}
	\label{tab:sim_res_samehaz}
\end{table}

\begin{table}[htbp]
	\centering\fontsize{9.5pt}{10pt}\selectfont
	\begin{tabular}{llcccccc}
		\toprule
		\multicolumn{2}{c}{ } & \multicolumn{3}{c}{\textbf{Intercept}} & \multicolumn{3}{c}{\textbf{Linear trend}} \\
		\cmidrule(l{3pt}r{3pt}){3-5} \cmidrule(l{3pt}r{3pt}){6-8}
		&  & \textbf{Bias} & \textbf{Emp. Var.} & \textbf{Covg.} & \textbf{Bias} & \textbf{Emp. Var.} & \textbf{Covg.}\\
		\midrule
		\addlinespace[0.3em]
		\multicolumn{6}{l}{\textbf{Vaccine efficacy constant at 75\%}}\\
		\addlinespace[0.3em]\hspace{1em}Cross at 150 cases & Constant VE & -0.012 & 0.039 & 0.951 & --- & --- & ---\\
		\hspace{1em} & log-linear & -0.013 & 0.046 & 0.953 & 0.002 & 0.111 & 0.952\\
		\hspace{1em} & P-spline & -0.018 & 0.076 & 0.970 & 0.002 & 0.109 & 0.954\\
		\addlinespace[0.3em]\hspace{1em}Cross at 1 year & Constant VE & -0.007 & 0.027 & 0.949 & --- & --- & ---\\
		\hspace{1em} & log-linear & -0.008 & 0.033 & 0.952 & -0.002 & 0.042 & 0.953\\
		\hspace{1em} & P-spline & -0.015 & 0.079 & 0.977 & -0.002 & 0.042 & 0.956\\
		\addlinespace[0.3em]\hspace{1em}Parallel trial & Constant VE & -0.004 & 0.013 & 0.951 & --- & --- & ---\\
		\hspace{1em} & log-linear & -0.007 & 0.046 & 0.950 & 0.000 & 0.040 & 0.950\\
		\hspace{1em} & P-spline & -0.023 & 0.110 & 0.977 & 0.002 & 0.039 & 0.952\\
		\bottomrule\\
		\addlinespace[0.3em]
		\multicolumn{8}{l}{\textbf{Vaccine efficacy wanes from 85\% to 35\% over 1.5 years}}\\
		\addlinespace[0.3em]\hspace{1em}Cross at 150 cases & log-linear & -0.016 & 0.050 & 0.951 & 0.002 & 0.064 & 0.952\\
		\hspace{1em} & P-spline & -0.025 & 0.086 & 0.975 & -0.002 & 0.063 & 0.953\\
		\addlinespace[0.3em]\hspace{1em}Cross at 1 year & log-linear & -0.010 & 0.036 & 0.952 & 0.003 & 0.031 & 0.953\\
		\hspace{1em} & P-spline & -0.022 & 0.085 & 0.979 & -0.004 & 0.031 & 0.955\\
		\addlinespace[0.3em]\hspace{1em}Parallel trial & log-linear & -0.011 & 0.049 & 0.950 & 0.007 & 0.031 & 0.948\\
		\hspace{1em} & P-spline & -0.030 & 0.124 & 0.980 & 0.003 & 0.031 & 0.949\\
		\bottomrule\\\vspace{0.1em}
	\end{tabular} 
	\caption{Bias, empirical variance, and coverage for estimates of the intercept and linear trend in vaccine efficacy under the log--linear model, (\ref{eqn:loglin_decay_haz}), and semi--parametric model, (\ref{eqn:pspline_haz}). Here, the time--varying  baseline hazard in year two the same as the baseline hazard in year one.}
	\label{tab:simres_par_ests_samehaz}
\end{table}

\newpage
\subsection{Comparing Uniform Crossover, Crossover at One Year, and Parallel Trials}

This section contains results for a set of idealized trials with constant baseline hazard, instantaneous enrollment and crossover, and constant VE. Trials either crossed placebo participants to the vaccine arm at one year, uniformly over the two year study period, or never (corresponding to a standard parallel arm design).  
Table 
\ref{tab:sim_res_idealized}  provides the results for the vaccine
efficacy over time while
Table
\ref{tab:par_ests_idealized} provides the parameter estimates.

\begin{table}[htbp]
	\centering\fontsize{9.5pt}{10pt}\selectfont
	\begin{tabular}{lccccccc}
		\toprule
		\multicolumn{2}{c}{ } & \multicolumn{3}{c}{$\boldsymbol{\log}\mathbf{(1 - VE(s))}$} & \multicolumn{3}{c}{$\boldsymbol{\log}\mathbf{(1 - VE(s))} - \boldsymbol{\log}\mathbf{(1 - VE(0))}$} \\
		\cmidrule(l{3pt}r{3pt}){3-5} \cmidrule(l{3pt}r{3pt}){6-8}
		\hspace{1em}\textbf{Model} & \textbf{Time} & \textbf{Bias} & \textbf{Empir. Var.} & \textbf{Coverage} & \textbf{Bias} & \textbf{Empir. Var.} & \textbf{Coverage}\\
		\midrule
		\addlinespace[0.3em]
		\multicolumn{8}{l}{\textbf{Continuous uniform crossover}}\\
		\cmidrule(lr){1-8}\hspace{1em} log-linear & 0.0 & -0.008 & 0.031 & 0.950 & --- & --- & ---\\
		\hspace{1em}  & 0.5 & -0.004 & 0.016 & 0.951 & 0.003 & 0.009 & 0.950\\
		\hspace{1em}  & 1.0 & -0.001 & 0.018 & 0.949 & 0.007 & 0.034 & 0.950\\
		\hspace{1em}  &1.5 & 0.002 & 0.037 & 0.951 & 0.010 & 0.077 & 0.950\\
		\hspace{1em}  & 2.0 & 0.006 & 0.074 & 0.950 & 0.013 & 0.137 & 0.950\\
		\addlinespace[0.3em]\hspace{1em}P-spline & 0.0 &  -0.017 & 0.072 & 0.979 & --- & --- & ---\\
		\hspace{1em}  & 0.5 & -0.006 & 0.023 & 0.981 & 0.010 & 0.082 & 0.983\\
		\hspace{1em}  & 1.0 & -0.004 & 0.029 & 0.976 & 0.012 & 0.104 & 0.967\\
		\hspace{1em}  & 1.5 & 0.002 & 0.044 & 0.973 & 0.018 & 0.103 & 0.970\\
		\hspace{1em}  & 2.0 & 0.011 & 0.183 & 0.976 & 0.027 & 0.234 & 0.980\\
		\bottomrule\\\addlinespace[0.3em]
		\multicolumn{8}{l}{\textbf{Crossover at one year}}\\
		\cmidrule(lr){1-8}\hspace{1em}log-linear & 0.0 &  -0.008 & 0.035 & 0.953 & --- & --- & ---\\
		\hspace{1em}  & 0.5 & -0.008 & 0.026 & 0.954 & 0.000 & 0.010 & 0.947\\
		\hspace{1em}  & 1.0 & -0.007 & 0.037 & 0.947 & 0.000 & 0.039 & 0.947\\
		\hspace{1em}  & 1.5 & -0.007 & 0.068 & 0.947 & 0.000 & 0.088 & 0.947\\
		\hspace{1em}  & 2.0 & -0.007 & 0.117 & 0.946 & 0.000 & 0.156 & 0.947\\
		\addlinespace[0.3em]\hspace{1em}P-spline &0.0 &  -0.018 & 0.100 & 0.984 & --- & --- & ---\\
		\hspace{1em} & 0.5 &  -0.009 & 0.032 & 0.982 & 0.009 & 0.112 & 0.986\\
		\hspace{1em}  & 1.0 & -0.012 & 0.048 & 0.975 & 0.005 & 0.122 & 0.971\\
		\hspace{1em}  & 1.5 & -0.010 & 0.078 & 0.970 & 0.007 & 0.130 & 0.981\\
		\hspace{1em}  &  2.0 & -0.002 & 0.237 & 0.980 & 0.016 & 0.333 & 0.975\\
		\bottomrule\\
		\addlinespace[0.3em]
		\multicolumn{8}{l}{\textbf{Parallel trial}}\\
		\cmidrule(lr){1-8}\hspace{1em}log-linear & 0.0 &  -0.006 & 0.051 & 0.950 & --- & --- & ---\\
		\hspace{1em}  & 0.5 & -0.007 & 0.022 & 0.950 & 0.000 & 0.010 & 0.951\\
		\hspace{1em} & 1.0 & -0.007 & 0.013 & 0.949 & -0.001 & 0.039 & 0.951\\
		\hspace{1em} & 1.5 & -0.008 & 0.024 & 0.951 & -0.001 & 0.088 & 0.951\\
		\hspace{1em} & 2.0 & -0.008 & 0.054 & 0.953 & -0.002 & 0.157 & 0.951\\
		\addlinespace[0.3em]\hspace{1em}P-spline & 0.0 &  -0.021 & 0.131 & 0.981 & --- & --- & ---\\
		\hspace{1em} & 0.5 & -0.011 & 0.032 & 0.982 & 0.010 & 0.140 & 0.986\\
		\hspace{1em} & 1.0 & -0.010 & 0.033 & 0.981 & 0.010 & 0.186 & 0.968\\
		\hspace{1em} & 1.5 & -0.010 & 0.034 & 0.983 & 0.011 & 0.165 & 0.973\\
		\hspace{1em} & 2.0 & -0.024 & 0.133 & 0.982 & -0.004 & 0.261 & 0.977\\
		\bottomrule\\
	\end{tabular}
	\caption{Bias, empirical variance, and coverage for the linear predictor in Cox PH models for simulated trials in an idealized scenario with constant baseline hazards and either continuous crossover, instantaneous crossover at one year, or a standard trial. The log-linear and P-spline models correspond to (\ref{eqn:loglin_decay_haz}) and (\ref{eqn:pspline_haz}), respectively.}
	\label{tab:sim_res_idealized}
\end{table}

\begin{table}[htbp]
	\centering \fontsize{9.5pt}{10pt}\selectfont
	\begin{tabular}{lcccccc}
		\toprule
		& \multicolumn{3}{c}{\textbf{Intercept}} & \multicolumn{3}{c}{\textbf{Linear trend}} \\
		\cmidrule(l{3pt}r{3pt}){2-4} \cmidrule(l{3pt}r{3pt}){5-7}
		& \textbf{Bias} & \textbf{Emp. Var.} & \textbf{Covg.} & \textbf{Bias} & \textbf{Emp. Var.} & \textbf{Covg.}\\
		\midrule
		\addlinespace[0.3em]
		\multicolumn{7}{l}{\textbf{Continuous uniform crossover}}\\
		\cmidrule(lr){1-7}\hspace{1em}log-linear & -0.008 & 0.031 & 0.950 & 0.007 & 0.034 & 0.950\\
		\hspace{1em}P-spline & -0.016 & 0.071 & 0.973 & 0.003 & 0.034 & 0.952\\
		\addlinespace[0.3em]
		\multicolumn{7}{l}{\textbf{Crossover at one year}}\\
		\cmidrule(lr){1-7}\hspace{1em}log-linear & -0.008 & 0.035 & 0.953 & 0.000 & 0.039 & 0.947\\
		\hspace{1em}P-spline & -0.018 & 0.099 & 0.978 & -0.002 & 0.038 & 0.952\\
		\addlinespace[0.3em]
		\multicolumn{7}{l}{\textbf{Parallel trial}}\\
		\cmidrule(lr){1-7}\hspace{1em}log-linear & -0.006 & 0.051 & 0.950 & -0.001 & 0.039 & 0.951\\
		\hspace{1em}P-spline & -0.021 & 0.130 & 0.977 & -0.001 & 0.039 & 0.952\\
		\bottomrule\\
	\end{tabular}
	\caption{Empirical variance and coverage for estimates of the intercept and linear trend in vaccine efficacy under the log--linear model, (\ref{eqn:loglin_decay_haz}), and semi--parametric model, (\ref{eqn:pspline_haz}), in an idealized scenario with constant baseline hazards  and 
		continuous crossover, instantaneous crossover at one year, or a standard trial.}
	\label{tab:par_ests_idealized}
\end{table}

\newpage
\subsection{Frailty simulation results}
\label{subsec:frailty_res}
This section presents results from simulated trials in where participants were heterogeneous in their baseline hazards. Simulation was analogous to trials simulated elsewhere in this manuscript, except that each trial consisted of 30,000 participants and the participant level hazard was $\widetilde{h}_i(t) = U_ih_i(t)$, with $h_i(t)$ corresponding to either a constant VE or log-linear VE model. In both cases the baseline hazard was constant. We considered two settings for the baseline hazard: a low event rate scenario calibrated to yield 100 cases per year on placebo or a high event rate scenario calibrated to yield 600 cases per year on placebo. Participant frailties were drawn from a gamma distribution with mean one and a variance of either one (low frailty variance scenario) or four (high frailty variance scenario). 

Tables 
\ref{tab:frailty_sumstats},
\ref{tab:ve_frailty_const},  and
\ref{tab:ve_frailty_waning}
present the results.

\begin{table}[htbp]
	\centering\fontsize{9.5pt}{10pt}\selectfont
	\begin{tabular}{ccccrrrrr}
		\toprule
		&&&&\multicolumn{5}{c}{\textbf{Frailty distribution summary statistics}}\\
		\cmidrule(lr){5-9}\shortstack{\textbf{Baseline}\\\textbf{Hazard}} & \shortstack{\textbf{Frailty}\\\textbf{Variance}} & \textbf{Design} & \textbf{Original arm} & \textbf{Mean} & \textbf{SD} & \textbf{25\%ile} & \textbf{50\%ile} & \textbf{75\%ile}\\
		\midrule
		\addlinespace[0.3em]
		\multicolumn{9}{l}{\textbf{VE constant at 75\%}}\\
		\cmidrule(lr){1-9}Low & Low & Cross at 1 year & Placebo & 0.992 & 0.992 & 0.285 & 0.688 & \vphantom{1} 1.375\\
		\hspace{1em} &  &  & Vaccine & 0.997 & 0.996 & 0.287 & 0.691 & \vphantom{1} 1.382\\
		\addlinespace[0.3em]\hspace{1em} &  & Parallel trial & Placebo & 0.987 & 0.987 & 0.284 & 0.684 & \vphantom{1} 1.368\\
		\hspace{1em} &  &  & Vaccine & 0.997 & 0.996 & 0.287 & 0.691 & 1.382\\
		\hspace{1em} & High & Cross at 1 year & Placebo & 0.969 & 1.937 & 0.010 & 0.169 & 1.010\\
		\hspace{1em} &  &  & Vaccine & 0.987 & 1.973 & 0.010 & 0.172 & \vphantom{1} 1.028\\
		\addlinespace[0.3em]\hspace{1em} &  & Parallel trial & Placebo & 0.949 & 1.897 & 0.010 & 0.166 & \vphantom{1} 0.989\\
		\hspace{1em} &  &  & Vaccine & 0.987 & 1.973 & 0.010 & 0.172 & 1.028\\
		\cmidrule(lr){1-9}High & Low & Cross at 1 year & Placebo & 0.955 & 0.955 & 0.275 & 0.662 & 1.323\\
		\hspace{1em} &  &  & Vaccine & 0.980 & 0.980 & 0.282 & 0.680 & \vphantom{1} 1.359\\
		\addlinespace[0.3em]\hspace{1em} &  & Parallel trial & Placebo & 0.926 & 0.926 & 0.266 & 0.642 & \vphantom{1} 1.283\\
		\hspace{1em} &  &  & Vaccine & 0.980 & 0.980 & 0.282 & 0.680 & 1.359\\
		\addlinespace[0.3em]\hspace{1em} & High & Cross at 1 year & Placebo & 0.841 & 1.681 & 0.009 & 0.147 & 0.876\\
		\hspace{1em} &  &  & Vaccine & 0.926 & 1.851 & 0.010 & 0.162 & \vphantom{1} 0.965\\
		\addlinespace[0.3em]\hspace{1em} &  & Parallel trial & Placebo & 0.757 & 1.514 & 0.008 & 0.132 & \vphantom{1} 0.790\\
		\hspace{1em} &  &  & Vaccine & 0.926 & 1.851 & 0.010 & 0.162 & 0.965\\
		\cmidrule(lr){1-7}\addlinespace[0.3em]
		\multicolumn{9}{l}{\textbf{VE wanes from 85\% to 35\% over 1.5 years}}\\
		\cmidrule(lr){1-9}\addlinespace[0.3em]Low & Low & Cross at 1 year & Placebo & 0.992 & 0.992 & 0.285 & 0.688 & 1.375\\
		\hspace{1em} &  &  & Vaccine & 0.994 & 0.994 & 0.286 & 0.689 & \vphantom{1} 1.378\\
		\addlinespace[0.3em]\hspace{1em} &  & Parallel trial & Placebo & 0.987 & 0.987 & 0.284 & 0.684 & 1.368\\
		\hspace{1em} &  &  & Vaccine & 0.994 & 0.994 & 0.286 & 0.689 & 1.378\\
		\addlinespace[0.3em]\hspace{1em} & High & Cross at 1 year & Placebo & 0.968 & 1.936 & 0.010 & 0.169 & 1.009\\
		\hspace{1em} &  &  & Vaccine & 0.975 & 1.950 & 0.010 & 0.170 & \vphantom{1} 1.017\\
		\addlinespace[0.3em]\hspace{1em} &  & Parallel trial & Placebo & 0.949 & 1.897 & 0.010 & 0.166 & 0.989\\
		\hspace{1em} &  &  & Vaccine & 0.975 & 1.950 & 0.010 & 0.170 & 1.017\\
		\cmidrule(lr){1-9}\addlinespace[0.3em]High & Low & Cross at 1 year & Placebo & 0.954 & 0.954 & 0.274 & 0.661 & 1.322\\
		\hspace{1em} &  &  & Vaccine & 0.964 & 0.964 & 0.277 & 0.668 & \vphantom{1} 1.337\\
		\addlinespace[0.3em]\hspace{1em} &  & Parallel trial & Placebo & 0.926 & 0.926 & 0.266 & 0.642 & 1.283\\
		\hspace{1em} &  &  & Vaccine & 0.964 & 0.964 & 0.277 & 0.668 & 1.337\\
		\addlinespace[0.3em]\hspace{1em} & High & Cross at 1 year & Placebo & 0.838 & 1.676 & 0.009 & 0.146 & 0.874\\
		\hspace{1em} &  &  & Vaccine & 0.870 & 1.740 & 0.009 & 0.152 & \vphantom{1} 0.907\\
		\addlinespace[0.3em]\hspace{1em} &  & Parallel trial & Placebo & 0.757 & 1.514 & 0.008 & 0.132 & 0.790\\
		\hspace{1em} &  &  & Vaccine & 0.870 & 1.740 & 0.009 & 0.152 & 0.907\\
		\bottomrule\vspace{1em}
	\end{tabular}
	\caption{Summary statistics of the frailty distribution of participants still in the risk set at the end of two years of followup. We report geometric means of summary statistics of each frailty distribution from 10,000 simulated trials.}
	\label{tab:frailty_sumstats}
\end{table}

\begin{table}[htbp]
	\centering\fontsize{9.5pt}{10pt}\selectfont
	\begin{tabular}{ccccccc}
		\toprule
		\multicolumn{3}{c}{ } & \multicolumn{2}{c}{\textbf{Placebo crossover}} & \multicolumn{2}{c}{\textbf{Parallel trial}} \\
		\cmidrule(l{3pt}r{3pt}){4-5} \cmidrule(l{3pt}r{3pt}){6-7}
		\shortstack{\textbf{Frailty}\\\textbf{variance}} & \textbf{Model} & \textbf{Time} & $\boldsymbol{\log}\mathbf{(1 -VE(s))}$ & \shortstack{$\boldsymbol{\log}\mathbf{(1 -VE(s))} - $\\\hspace{0.1in}$\boldsymbol{\log}\mathbf{(1 -VE(0))}$} & $\boldsymbol{\log}\mathbf{(1 -VE(s))}$ & \shortstack{$\boldsymbol{\log}\mathbf{(1 -VE(s))} - $\\\hspace{0.1in}$\boldsymbol{\log}\mathbf{(1 -VE(0))}$}  \\
		\midrule
		\addlinespace[0.3em]
		\multicolumn{7}{l}{\textbf{Low baseline hazard}}\\
		\cmidrule(lr){1-7}Low & log-linear & 0.5 & -0.014 & 0.000 & -0.013 & 0.001\\
		\hspace{1em} &  & 1.0 & -0.015 & 0.000 & -0.012 & 0.002\\
		\hspace{1em} &  & 1.5 & -0.015 & -0.001 & -0.011 & 0.003\\
		\hspace{1em} &  & 2.0 & -0.015 & -0.001 & -0.010 & 0.005\\
		\addlinespace[0.3em]\hspace{1em} & P-spline & 0.5 & -0.017 & 0.022 & -0.017 & 0.043\\
		\hspace{1em} &  & 1.0 & -0.022 & 0.016 & -0.017 & 0.043\\
		\hspace{1em} &  & 1.5 & -0.020 & 0.018 & -0.019 & 0.040\\
		\hspace{1em} &  & 2.0 & -0.032 & 0.006 & -0.046 & 0.014\\
		\cmidrule(lr){1-7}\addlinespace[0.3em]High & log-linear & 0.5 & -0.009 & 0.009 & -0.007 & 0.011\\
		\hspace{1em} &  & 1.0 & 0.000 & 0.019 & 0.004 & 0.022\\
		\hspace{1em} &  & 1.5 & 0.010 & 0.028 & 0.015 & 0.032\\
		\hspace{1em} &  & 2.0 & 0.019 & 0.038 & 0.026 & 0.043\\
		\addlinespace[0.3em]\hspace{1em} & P-spline & 0.5 & -0.014 & 0.021 & -0.014 & 0.041\\
		\hspace{1em} &  & 1.0 & -0.007 & 0.028 & 0.001 & 0.057\\
		\hspace{1em} &  & 1.5 & 0.003 & 0.038 & 0.008 & 0.063\\
		\hspace{1em} &  & 2.0 & 0.005 & 0.041 & -0.017 & 0.039\\
		\cmidrule(lr){1-7}\addlinespace[0.3em]
		\multicolumn{7}{l}{\textbf{High baseline hazard}}\\
		\cmidrule(lr){1-7}\addlinespace[0.3em]Low & log-linear & 0.5 & 0.011 & 0.016 & 0.012 & 0.015\\
		\hspace{1em} &  & 1.0 & 0.026 & 0.031 & 0.027 & 0.029\\
		\hspace{1em} &  & 1.5 & 0.042 & 0.047 & 0.041 & 0.044\\
		\hspace{1em} &  & 2.0 & 0.058 & 0.063 & 0.056 & 0.059\\
		\addlinespace[0.3em]\hspace{1em} & P-spline & 0.5 & 0.010 & 0.017 & 0.010 & 0.018\\
		\hspace{1em} &  & 1.0 & 0.026 & 0.033 & 0.027 & 0.036\\
		\hspace{1em} &  & 1.5 & 0.042 & 0.050 & 0.042 & 0.050\\
		\hspace{1em} &  & 2.0 & 0.052 & 0.059 & 0.046 & 0.054\\
		\cmidrule(lr){1-7}\addlinespace[0.3em]High & log-linear & 0.5 & 0.054 & 0.054 & 0.055 & 0.050\\
		\hspace{1em} &  & 1.0 & 0.108 & 0.108 & 0.105 & 0.100\\
		\hspace{1em} &  & 1.5 & 0.162 & 0.162 & 0.154 & 0.149\\
		\hspace{1em} &  & 2.0 & 0.216 & 0.216 & 0.204 & 0.199\\
		\addlinespace[0.3em]\hspace{1em} & P-spline & 0.5 & 0.052 & 0.054 & 0.053 & 0.057\\
		\hspace{1em} &  & 1.0 & 0.108 & 0.109 & 0.108 & 0.112\\
		\hspace{1em} &  & 1.5 & 0.162 & 0.164 & 0.154 & 0.159\\
		\hspace{1em} &  & 2.0 & 0.210 & 0.211 & 0.190 & 0.194\\
		\bottomrule\vspace{0.3em}
	\end{tabular}
	\caption{Bias of estimates of the linear predictor for VE and change in the linear predictor for VE in Cox PH models for trials simulated with constant VE at 75\% and gamma distributed frailties. The low baseline hazard scenario was calibrated to yield an average of 50 cases per six month period on the placebo arm, while the high baseline hazard scenario was calibrated to yield 300 cases per six month period. The frailty distribution had mean one and a variance of either one (low variance) or four (high variance).}
	\label{tab:ve_frailty_const}
\end{table}

\begin{table}[htbp]
	\centering\fontsize{9.5pt}{10pt}\selectfont
	\begin{tabular}{ccccccc}
		\toprule
		\multicolumn{3}{c}{ } & \multicolumn{2}{c}{\textbf{Placebo crossover}} & \multicolumn{2}{c}{\textbf{Parallel trial}} \\
		\cmidrule(l{3pt}r{3pt}){4-5} \cmidrule(l{3pt}r{3pt}){6-7}
		\shortstack{\textbf{Frailty}\\\textbf{variance}} & \textbf{Model} & \textbf{Time} & $\boldsymbol{\log}\mathbf{(1 -VE(s))}$ & \shortstack{$\boldsymbol{\log}\mathbf{(1 -VE(s))} - $\\\hspace{0.1in}$\boldsymbol{\log}\mathbf{(1 -VE(0))}$} & $\boldsymbol{\log}\mathbf{(1 -VE(s))}$ & \shortstack{$\boldsymbol{\log}\mathbf{(1 -VE(s))}-  $\\\hspace{0.1in}$\boldsymbol{\log}\mathbf{(1 -VE(0))}$} \\
		\midrule
		\addlinespace[0.3em]
		\multicolumn{7}{l}{\textbf{Low baseline hazard}}\\
		\cmidrule(lr){1-7}\addlinespace[0.3em]Low & log-linear & 0.5 & -0.016 & 0.005 & -0.013 & 0.008\\
		\hspace{1em} &  & 1.0 & -0.010 & 0.011 & -0.005 & 0.015\\
		\hspace{1em} &  & 1.5 & -0.005 & 0.016 & 0.002 & 0.023\\
		\hspace{1em} &  & 2.0 & 0.001 & 0.022 & 0.010 & 0.030\\    
		\addlinespace[0.3em]\hspace{1em} & P-spline & 0.5 & -0.019 & 0.036 & -0.020 & 0.058\\
		\hspace{1em} &  & 1.0 & -0.010 & 0.045 & -0.006 & 0.071\\
		\hspace{1em} &  & 1.5 & -0.001 & 0.054 & 0.000 & 0.077\\
		\hspace{1em} &  & 2.0 & 0.009 & 0.064 & 0.005 & 0.083\\
		\cmidrule(lr){1-7}\addlinespace[0.3em]High & log-linear & 0.5 & -0.008 & 0.011 & -0.004 & 0.013\\
		\hspace{1em} &  & 1.0 & 0.003 & 0.022 & 0.009 & 0.025\\
		\hspace{1em} &  & 1.5 & 0.014 & 0.033 & 0.022 & 0.038\\
		\hspace{1em} &  & 2.0 & 0.026 & 0.044 & 0.034 & 0.051\\
		\addlinespace[0.3em]\hspace{1em} & P-spline & 0.5 & -0.012 & 0.037 & -0.014 & 0.059\\
		\hspace{1em} &  & 1.0 & 0.005 & 0.054 & 0.011 & 0.083\\
		\hspace{1em} &  & 1.5 & 0.020 & 0.069 & 0.022 & 0.095\\
		\hspace{1em} &  & 2.0 & 0.030 & 0.079 & 0.022 & 0.095\\
		\addlinespace[0.3em]
		\multicolumn{7}{l}{\textbf{High baseline hazard}}\\
		\cmidrule(lr){1-7}\addlinespace[0.3em]Low & log-linear & 0.5 & 0.011 & 0.012 & 0.015 & 0.010\\
		\hspace{1em} &  & 1.0 & 0.023 & 0.023 & 0.025 & 0.021\\
		\hspace{1em} &  & 1.5 & 0.035 & 0.035 & 0.035 & 0.031\\
		\hspace{1em} &  & 2.0 & 0.046 & 0.047 & 0.046 & 0.041\\
		\addlinespace[0.3em]\hspace{1em} & P-spline & 0.5 & 0.011 & 0.021 & 0.011 & 0.022\\
		\hspace{1em} &  & 1.0 & 0.027 & 0.037 & 0.029 & 0.040\\
		\hspace{1em} &  & 1.5 & 0.038 & 0.048 & 0.038 & 0.049\\
		\hspace{1em} &  & 2.0 & 0.041 & 0.051 & 0.038 & 0.049\\
		\cmidrule(lr){1-7}\addlinespace[0.3em]High & log-linear & 0.5 & 0.056 & 0.038 & 0.063 & 0.033\\
		\hspace{1em} &  & 1.0 & 0.095 & 0.077 & 0.096 & 0.066\\
		\hspace{1em} &  & 1.5 & 0.133 & 0.115 & 0.128 & 0.099\\
		\hspace{1em} &  & 2.0 & 0.172 & 0.154 & 0.161 & 0.131\\
		\addlinespace[0.3em]\hspace{1em} & P-spline & 0.5 & 0.056 & 0.059 & 0.057 & 0.063\\
		\hspace{1em} &  & 1.0 & 0.108 & 0.110 & 0.109 & 0.114\\
		\hspace{1em} &  & 1.5 & 0.142 & 0.145 & 0.135 & 0.141\\
		\hspace{1em} &  & 2.0 & 0.151 & 0.153 & 0.136 & 0.141\\
		\bottomrule\vspace{0.3em}
	\end{tabular}
	\caption{Bias of estimates of the linear predictor for VE and change in the linear predictor for VE in Cox PH models for for trials simulated with VE waning from 85\% to 35\% linear on the log hazard scale and gamma distributed frailties. The low baseline hazard scenario was calibrated to yield an average of 50 cases per six month period on the placebo arm, while the high baseline hazard scenario was calibrated to yield 300 cases per six month period. The frailty distribution had mean one and a variance of either one (low variance) or four (high variance).}
	\label{tab:ve_frailty_waning}
\end{table}

\end{document}